\documentclass[11pt,a4paper,nofootinbib]{article}
\usepackage[english]{babel}
\usepackage[isolatin]{inputenc} 

\usepackage{etaremune}
\usepackage{amsfonts,amssymb,amsmath,bbm}
\usepackage{fontenc,indentfirst, delarray}

\usepackage[english]{babel}
\usepackage[isolatin]{inputenc} 

\usepackage{etaremune}
\usepackage{amsfonts,amssymb,amsmath,bbm}
\usepackage{fontenc,indentfirst, delarray}
\usepackage{graphicx,graphics,xcolor,epstopdf,epsf}
\DeclareGraphicsExtensions{eps,ps,p.gz,pdf}
\usepackage{pdfsync}
\usepackage[pdfstartview=FitH]{hyperref}
\usepackage{setspace}

\usepackage{footmisc}
\usepackage{fancyhdr}

\makeatletter
\renewcommand{\section}{\@startsection {section}{1}{\z@}%
             {-3.5ex \@plus -1ex \@minus -.2ex}%
             {2.3ex \@plus.2ex}%
             {\normalfont\normalsize\sffamily\bfseries}}
\renewcommand{\subsection}{\@startsection {subsection}{1}{\z@}%
             {-3.5ex \@plus -1ex \@minus -.2ex}%
             {2.3ex \@plus.2ex}%
             {\normalfont\normalsize\sffamily\emph}}

\@addtoreset{equation}{section}

\makeatother

\newtheorem{thm}{Theorem}[section]
\newtheorem{lem}[thm]{Lemma}
\newtheorem{prop}[thm]{Proposition}

\newtheorem{defi}[thm]{Definition}
\newtheorem{rem}[thm]{Remark}
\newenvironment{preuve}{{\emph{Proof.}}}{\hfill$\blacksquare$}

\newcommand{\norm}[1]{\left\lVert#1\right\rVert}
\newcommand{\abs}[1]{\lvert#1\rvert}

\newcommand{\scl}[2]{\langle#1,#2\rangle}
\newcommand{\suup}[1]{ \underset{#1}{\sup} }

\newcommand{\ket}[1]{\lvert#1\rangle}

\def\begf{\begin{frame}}
\def\enf{\end{frame}}

\def\begz{\begin{itemize}}

\def\endz{\end{itemize}}

\def\lp{\left(} 
\def\rp{\right)} 
\def\dm{\lp\begin{array}}	
\def\fm{\end{array}\rp}
\def\m2{M_2 \lp \cc \rp}
\def\m3{M_3 \lp \cc \rp}

\def\ot{\otimes}

\def\ds{\partial\!\!\!\slash}

\def\kk{{\mathbb{K}}}	
\def\cc{{\mathbb{C}}}
\def\C{{\mathbb{C}}}	

\def\R{{\mathbb{R}}}

\def\N{{\mathbb{N}}}
 		
\def\ii{{\mathbb{I}}}
\def\I{{\mathbb{I}}}
\def\hhh{{\mathbb H}}

\def\mm{{\mathcal M}}	
\def\M{{\mathcal M}}		
\def\aa{{\mathcal A}}

\def\A{{\mathcal A}}			
\def\bb{{\mathcal B}}			
\def\ll{{\mathcal L}}
\def\dd{{\mathcal D}}	
\def\hh{{\mathcal H}}

\def\pp{{\mathcal P}}
\def\ccc{{\mathcal C}}

\def\ss{{\mathcal S}}

\def\coi{C^{\infty}_0\lp\mm\rp}

\def\ot{\otimes}

\def\xo0{\omega^0_x}
\def\yo0{\omega^0_y}

\def\xo0{x_\omega^0}
\def\yo0{y_\omega^0}

\def\pa{{\cal P}(\aa)}
\def\sa{{\cal S}(\aa)}
\def\qq{{\cal Q}}

\def\fm{\Phi(x^\mu)}

\def\dm{\partial_\mu}

\def\dmm{\left(\begin{array}}
\def\fmm{\end{array}\right)}

\newcommand{\HH}{\mathcal{H}}

\def\s0a{{\cal S}_0(\A)}

\def\j0{{\bf J_0}}

\begin{document}
\thispagestyle{empty}

\title{\vspace{-2truecm}Noncommutative geometry of the Moyal plane: \\
translation isometries, Connes' distance on
  coherent states,\\ Pythagoras equality. \footnotetext{\hspace{-.65truecm}Work supported by the {\bf ERC Advanced Grant} 227458 OACFT
  \emph{Operator Algebras \!\&\! Conformal Field Theory} and the
  {\bf ERG-Marie Curie fellowship} 237927 \emph{Noncommutative geometry \!\&\!
    quantum gravity}.}}
 \date{\today}

 \author{Pierre
   Martinetti$^{a,b,c}${\footnote{Pierre.Martinetti@roma1.infn.it}}
\, ,\,
Luca Tomassini$^{a}${\footnote{tomassin@mat.uniroma2.it}}
}
\date{
$^a$ CMTP \& Dipartimento di Matematica,~Universit\`a di Roma Tor Vergata, I-00133;\\ 
$^b$ Dipartimento di Fisica,~Universit\`a di Roma ``Sapienza'',
I-00185;\\
$^c$ Dipartimento di Fisica,~Universit\`a di Napoli Federico II, I-00185.}
\maketitle
\abstract{We study the metric aspect of the Moyal plane
from Connes' noncommutative geometry point of view. First, we compute Connes' spectral
  distance associated with the natural isometric action of $\R^2$ on the algebra
  of the Moyal plane $\A$. We show that the distance between any state
  of $\A$ and
  any of its translated is precisely the amplitude of
  the translation. As a consequence, we obtain the spectral distance between coherent states of the quantum
  harmonic oscillator as the Euclidean distance on the
  plane. We investigate the classical limit,
showing that the set of coherent states equipped with Connes' spectral
distance tends towards the Euclidean plane as the parameter of
deformation goes to zero.  The extension of these results to the action of
the symplectic group is also discussed, with particular emphasize on
the orbits of coherent states under rotations. Second, we compute the spectral
  distance in the double Moyal plane, intended as the product of (the
  minimal unitization of) $\A$ by $\C^2$. We show that on the set of states obtained by translation of an arbitrary state of $\A$, this distance is given by Pythagoras theorem. On the way, we prove some Pythagoras
inequalities for the product of arbitrary unital \& non-degenerate
spectral triples. 
Applied to the
  Doplicher-Fredenhagen-Roberts model of quantum spacetime  [DFR],
  these two theorems show that Connes'
spectral distance and the DFR quantum length coincide on the set of states
of optimal localization. 

\tableofcontents

\section{Introduction}

Long after their introduction for the study of quantum mechanics in phase
space \cite{Groene, Moyal}, Moyal spaces are now
intensively used in physics and mathematics as a paradigmatic example of
noncommutative geometry by deformation (especially, in most recent time, with the aim of developing
quantum field theory on noncommutative spacetime). However, their metric
aspect has been little studied. The direct approach, consisting in
deforming the Riemannian metric tensor by lifting the star
product \cite{madore}, does not allow to construct a ``noncommutative'' line element  that would be integrated along a ``Moyal-geodesic'' in order to get a
``quantum distance''. Nevertheless, there exist (at least) two alternative proposals for
extracting some metric information
from Moyal spaces, both starting with an algebraic formulation of the distance: one is Connes'
spectral distance formula \cite{Connes:1994kx}, the other is the length operator in the
Doplicher-Fredenhagen-Roberts model of quantum spacetime [DFR]
\cite{Doplicher:1995hc}.  In this paper, we prove two theorems on the
spectral distance: the first one gives the distance between any two states
obtained from one another by an isometric action of $\R^2$ on the
Moyal plane, the second
is a Pythagoras equality in the double Moyal
plane. Besides
their own interest (few such explicit general results are known on the metric aspect of noncommutative
geometry), these two results 
allow us to show
\cite{Martinetti:2011fk} how the spectral distance and the DFR quantum
length, restricted to the set of physically relevant states, capture
the same metric information on a quantum space. 
\newline

Recall that, given a spectral triple \cite{Connes:1994kx} (or unbounded
Fredholm module) $T = (\A, \HH, D)$
where
\begin{enumerate}
\item[-] $\A$ is an involutive algebra acting by $\pi$ on a
  Hilbert space $\HH$;
\item[-]  the so called Dirac operator $D$ is a non-necessarily bounded,
densely defined,  selfadjoint operator on $\HH$, such that
$\pi(a)(D-\lambda\ii)^{-1}$ is compact for any $a\in\A$ and
$\lambda$ in the resolvent set of $D$ (in case $\A$ is unital, this means
$D$ has compact resolvent); 
\item[-]  the set $\{a\in \A,
[D, \pi(\A)]\in \bb(\HH)\}$ is dense in $\A$; 
\end{enumerate}
 Connes has proposed on the state space $\sa$ of  $\A$ (see
 \S Notations) the following
 distance \cite{Connes:1989fk},
\begin{equation}
  \label{eq:22}
  d_D(\varphi, \tilde\varphi) \doteq
  \sup_{a\in\bb_{\text{Lip}}(T)}\abs{\varphi (a) - \tilde\varphi(a)},
\end{equation}
where $\varphi,\tilde\varphi\in\sa$ are any two states and
\begin{equation}
  \label{eq:25}
  \bb_{\text{Lip}}(T) \doteq \left\{ a\in \A, \; \norm{[D,\pi(a)]}\leq 1\right\}
\end{equation}
denotes the $D$-Lipschitz ball of $\A$, that is the unit ball for the
Lipschitz semi-norm
\begin{equation}
  \label{eq:53}
  L(a) \doteq \norm{[D,\pi(a)]},
\end{equation}
where $\norm{.}$ is the operator norm
coming from the representation $\pi$.

 In case $\A=C^\infty(\M)$ is the (commutative) algebra of smooth functions
on a compact Riemannian spin
manifold $\M$, with $D=\ds\doteq -i\sum_\mu \gamma^\mu\partial_\mu$ the Dirac operator
of quantum field theory and $\HH$ the Hilbert space of square
integrable spinors on $\M$, the spectral distance $d_{\ds}$ coincides
with the Monge-Kantorovich (also called Wasserstein) distance of order $1$ in the theory of optimal transport
\cite{Rieffel:1999ec}.
This result still holds for locally compact
manifolds, as soon as they are geodesically
complete \cite{dAndrea:2009xr}.  For pure states, that is - by Gelfand
theorem -  
evaluation at points $x$ of $\M$ - $\delta_{x}(f) \doteq f(x)$ for
$f\in\coi$ - one retrieves the geodesic distance 
associated with the Riemannian structure,
\begin{equation}
\label{dgeo}
d_{\ds} (\delta_{x}, \delta_{y}) = d_{\text{geo}}(x, y).
\end{equation}

Therefore, the spectral distance appears as an alternative to the
usual definition of the geodesic distance, whose advantage is to 
make sense also in a
noncommutative context. It has been explicitly calculated in several noncommutative spectral triples inspired by high
energy physics \cite{Connes:1996fu}, providing an interpretation to
the recently discovered Higgs field \cite{Collaboration:2012fk} as the component of the
metric in a discrete internal dimension \cite{Connes:1996fu,Martinetti:2002ij},
and exhibiting intriguing links with other distances, like the
Carnot-Carath\'eodory metric in sub-riemannian geometry \cite{Martinetti:2006db,Martinetti:2008hl}.
Various examples with finite dimensional algebras have also been
investigated \cite{bimonte, pekin,dimakis2,Iochum:2001fv}, as well as
for fractals \cite{Christensen:2006fk, Christensen:2011fk} and the
noncommutative torus \cite{Cagnache:2009vn}.

As often advertised by Connes, formula
(\ref{eq:22}) is particularly interesting for it does not
rely on any notion ill-defined in a quantum context, such as points or
path between points. In this perspective, the spectral distance seems more
compatible with a (still unknown) description of spacetime at the Planck
scale than the distance  viewed as the length of the shortest
path.  To push this idea further, one investigated  in
\cite{Cagnache:2009oe} the spectral distance for the simplest spectral
triple one may associate to quantum mechanics, namely the isospectral
deformation of the Euclidean space based on the noncommutative Moyal
product $\star$
\cite{Gayral:2004rc}. For technical reasons, in \cite{Cagnache:2009oe} only the stationary 
states of the quantum harmonic oscillator were
taken into account. In the present
paper, we extend the analysis to a wider class of states, including
coherent states. 
\newline

Our first result is theorem
\ref{theomain}: the spectral distance between
\emph{any} state $\varphi$ of the $2$-dimensional Moyal algebra $\A$ and any of its
translated $\varphi_\kappa$, $\kappa\in\R^2$, is the
(geodesic) length of translation,
\begin{equation}
\label{eq:171}
d_D(\varphi, \varphi_\kappa) = \abs{\kappa}.
\end{equation}
The extension of this
result to the action of the symplectic group is discussed in
section \ref{symplectic} where we provide an upper bound to the spectral
distance on any symplectic orbit (proposition \ref{propsymp}).

The second result is a Pythagoras equality in the
double Moyal plane (theorem~\ref{theopyth}). By this,  we mean the
product $T'$ (intended as an ordered unit space, details are given in
due time) of the minimal unitalization $T^+$ of  the spectral triple of
the Moyal plane with the canonical spectral triple on $\C^2$. For a fixed state $\varphi$ of the Moyal algebra $\A$, we show the spectral distance $d_{D'}$ on the subset
of $\ss(\A\otimes\C^2)$ given by 
\begin{equation}
\left\{(\varphi_\kappa, \delta^i),  \kappa\in\R^2, i=1,2\right\}
\text{ with } \delta^1, \delta^2 \text{ the two pure states of }
\C^2, 
\label{eq:170}
\end{equation}
satisfies Pythagoras theorem, that is
\begin{equation}
  \label{eq:172}
  d^2_{D'} \left( \left(\varphi, \delta^1\right),\left(\varphi_\kappa,
      \delta^2\right)\right) = d^2_{D'} \left( \left(\varphi, \delta^1\right),
    \left(\varphi_\kappa, \delta^1\right)\right) +  d^2_{D'} \left( \left(\varphi_\kappa, \delta^1\right),
    \left(\varphi_\kappa, \delta^2\right)\right).
\end{equation}
Such an equality was known for the product of a manifold by
$\C^2$  \cite{Martinetti:2002ij} or - for a very particular class of states -  
for the product of a manifold by some finite dimensional
noncommutative algebra \cite{Martinetti:2001fk}.
The remarkable point here is
that Pythagoras theorem holds true for $\A$ an infinite dimensional noncommutative algebra.
As a side result, we also obtain in proposition
\ref{pyth} some
Pythagoras inequalities that hold true in full generality, meaning for
any states in the
product of any unital and non-degenerate spectral triple with~$\C^2$. 
 
These two theorems allow to show in section \ref{secDFR} that the spectral distance on the double
Moyal space coincides with the quantum length in the DFR model. 
By the first theorem, we also obtain in proposition \ref{propcoherent0} the
spectral distance $d_D$
between coherent states of the one dimensional  quantum harmonic
oscillator as the Euclidean distance on the plane. The classical limit
of the Moyal plane as a metric space is investigated in 
section \ref{coherentsection}.   

Although the paper is self-contained, part of it can
be thought as a continuation
  of \cite{Cagnache:2009oe},  as well as a companion to
  \cite{Martinetti:2011fk}. A non-technical presentation of part of
  these results can be found in \cite{Martinetti:2011fkbis}.
\newline

The paper is organized as follows. In section \ref{Moyalsection}}, we recall some basic properties of the Moyal
plane and its link with quantum mechanics. We emphasize
the unitary implementation of the translations, both in the
left-regular and the Schr\"odinger representations. Section
\ref{distancetranslationsection} contains the proof of
the first theorem, eq. (\ref{eq:171}). Using the characterization of the Lipschitz ball
in the Schr\"o\-dinger representation provided in section \ref{Moyalsection}, we show in section
\ref{secup} that $d_D(\varphi,\varphi_\kappa)\leq \abs{\kappa}$. The
most technical part of the proof consists in exhibiting a sequence of
elements in the Lipschitz ball that attains this upper bound. This is
done in sections \ref{optsection}
and \ref{subsecmain}. The
result is discussed in section \ref{remradial}, in the light of the
commutative case. Section \ref{symplectic} is
about the extension of these results to the symplectic group.  Section \ref{secPyth}
deals with the double Moyal space and Pythagoras theorem. We first
prove some Pythagoras inequalities for the product of an arbitrary
unital non-degenerate spectral triple with $\C^2$ (section~\ref{Pythineq}), then
Pythagoras equa\-lity for the Moyal plane in section \ref{pythagore}. Section
\ref{secdfr} deals with the application to the
coherent states, the classical
limit, and the DFR model. 
\newline

\noindent {\bf Notations and terminology:} 
Formula (\ref{eq:22}) has all the
properties of a
distance, except it might be
infinite. We should call it pseudo-distance but
for brevity we omit ``pseudo''.
For
coherence, we keep  the terminology used in
\cite{Martinetti:2008hl,dAndrea:2009xr,Martinetti:2011fk,Cagnache:2009oe}
and called $d_D$ the \emph{spectral distance},
warning the reader that - e.g. in
\cite{Bellissard:2010fk} -
formula (\ref{eq:22}) is called \emph{Connes distance} and is denoted $d_C$.

A state $\varphi$ of a $C^*$-algebras
is a positive ($\varphi(a^*a)\geq 0$) normalized
($\norm{\varphi}=1)$ linear form, with
\begin{equation}
  \label{eq:118}
\norm{\varphi}\doteq \suup{0\neq a\in\A}
\abs{\varphi(a)}\norm{a}^{-1}.  
\end{equation}
It is pure when  it cannot be written as a convex combination
of two other
states. The set of states of $\A$, respectively pure
states, is denoted $\sa$, resp. $\pa$. In case the algebra $\A$ in a
spectral triple $T$ is not
$C^*$, we call ``state''
the restriction $\varphi_0$ to 
$\A$ of a state $\varphi$ of the $C^*$-closure
of $\pi(\A)$. Then $\sa, \pa$ are shorthand notations for
$\ss(\overline{\pi(\A)})$, $\pp(\overline{\pi(\A)})$. By
continuity in the $C^*$-norm, $\varphi_0 = \tilde\varphi_0$ if and only if
$\varphi=\tilde\varphi$. So there is no use to distinguish between a
state and its restriction and we use the same symbol $\varphi$ for
both. 

Dirac bracket $\scl{\cdot}{\cdot}$  and parenthesis
$(\cdot,\cdot)$  denote the inner products  on  $L^2(\R)$ and
$L^2(\R^2)$. $\ii$ and $\ii_N$, $N\geq 2$,  are the
identity operators on the infinite and $N$-dimensional 
separable Hilbert spaces. Gothic letters $\mathfrak q, \mathfrak p...$
denote operators on $L^2(\R)$ (i.e. in the Schr\"odinger
representation). $S(\R^d)$ is the space of Schwartz functions on the
Euclidean space of dimension $d$.

The identity of a unital algebra is denoted ${\bf 1}$.
We use Einstein summation on alternate (up/down) indices. 
\newpage
\section{Moyal plane}
\label{Moyalsection}
We recall the definition of the spectral triple
associated to the Moyal space and stress the interest to switch from  the left-regular representation $\ll$ of the Moyal algebra
on $L^2(\R^{2N})$ to the (integrated)  Schr\"odinger
representation $\pi_S$ on $L^2(\R^N)$, in order to get an easy characterization of the
Lipschitz ball (lemma \ref{lipschro}). On our way, we collect various
formulas that will be useful for subsequent calculations, including
the unitary implementation of the translations in the Moyal plane. 
Most of this is very well known from von Neumann uniqueness
theorem. Nevertheless, it is useful  to have all this material,
sometimes a bit spread out in the literature, gathered in one single
section.  The reader familiar with Moyal quantization is invited to
jump to section III.
\subsection{Spectral triple for the Moyal plane}

Hereafter, we call Moyal algebra the noncommutative $\star$-deformation of the algebra of
Schwartz functions $S(\R^{2N})$ (with its standard Fréchet topology) by a non-degenerate symplectic form 
$\sigma$ on $\R^{2N}$ with determinant $\theta^{2N}\in (0,1]$,
\begin{equation}
\label{eq:24}
  (f\star g)(x) \doteq \frac 1{(\pi\theta)^{2N}}\int_{\R^{4N}} \,d^{2N}s
  \;d^{2N}t \; f(x+s) \,g(x+t) \,e^{-2i\sigma(s,t)}
\end{equation}
for $f,g\in S(\R^{2N})$, with
\begin{equation}
\sigma(s,t) = \frac 1{\theta} \, s^\mu \,\Theta_{\mu\nu}\, t^\nu,
\quad \,\Theta = \left(
\begin{array}{cc} 
0& -\I_N \\
 \I_N & 0 
\end{array}\right).
\label{eq:45}
\end{equation}

A so called \emph{isospectral deformation}
\cite{Connes:2002xr,sitarz} of the Euclidean space is a
spectral triple in which the algebra is a noncommutative deformation of some
commutative algebra of functions on the space,  while the Dirac operator keeps the same
spectrum as in the commutative case. For instance 
\begin{equation}
  \label{eq:2}
  \A = (S(\R^{2N}),\star),\; \hh = L^2(\R^{2N})\otimes \C^M,\, D= -i\gamma^\mu\partial\mu
\end{equation}
satisfy the properties of a spectral triple \cite{Gayral:2004rc,Wallet:2011uq}. Here
$M\doteq 2^{N}$ is the dimension of the spin repre\-sen\-tation, the
$\gamma^\mu$'s are the Euclidean Dirac matrices satisfying (with
$\delta^{\mu\nu}$ the Euclidean metric)
\begin{equation}
  \label{eq:28}
  \gamma^\mu\gamma^\nu + \gamma^\mu\gamma^\nu = 2\delta^{\mu\nu}\ii_M
  \quad \forall  \mu,\nu  = 1,...,2N,
\end{equation}
and the representation $\pi$ of $\A$ on $\HH$ is a multiple of the left regular representation
\begin{equation}
  \label{eq:5}
  \ll(f)\psi \doteq f\star \psi\quad\quad \forall f\in\A,\, \psi\in L^2(\R^{2N}),
\end{equation}
that is 
\begin{equation}
  \label{eq:23}
  \pi(f) \doteq  \ll(f)\otimes \ii_M.
\end{equation}

  In the following we restrict to the Moyal plane $N=1$, although the
extension of our results to arbitrary $N$ should be straightforward. So, from now on,
\begin{equation}
  \label{eq:42}
  \A =  (S(\R^{2}),\star).
\end{equation}
The plane $\R^2$ is parametrized by Cartesian coordinates
$x_\mu$ with derivative $\partial_\mu$, $\mu=1,2$. We denote
  \begin{equation}
\label{defcoorz}
 z \doteq\frac{x_1 + ix_2}{\sqrt 2},\quad \bar z \doteq \frac{x_1-ix_2}{\sqrt 2},
\end{equation}
with corresponding derivatives
\begin{equation}
  \label{eq:18}
  \partial \doteq \partial_z = \frac 1{\sqrt
    2}(\partial_1 - i \partial_2),\quad
  \bar\partial \doteq \partial_{\bar z} = \frac 1{\sqrt
2}(\partial_1  + i\partial_2).
\end{equation} 
The Dirac operator
\begin{equation}
  \label{eq:33}
  D = -i\sigma^\mu\partial_\mu = -i\sqrt 2\left(\begin{array}{cc} 0
      & \bar\partial \\ \partial & 0\end{array}\right),
\end{equation}
with $\sigma^\mu$ the Pauli matrices, acts as a first order differential operator on
\begin{equation}
\HH = L^2(\R^2)\otimes \C^2.
\label{eq:41}
\end{equation}
Notice that the spectral triple
\begin{equation}
T=(\A, \hh, D)\label{eq:184}
\end{equation}
of the Moyal
plane is non-unital ($\A$ has no unit) and
non-degenerate ($\pi(a)\psi = 0 \;\,\forall a \in\A$ implies
$\hh\ni \psi=0$). This point will be important when discussing
Pythagoras theorem.

The commutator of $D$ with a Schwartz function $f$ acts by $\star$-multiplication on
\begin{equation}
\label{eq:35}
\psi = \left(
\begin{array}{c} 
\psi_1 \\ 
\psi_2
\end{array}\right)
\in \HH,
\end{equation}
that is 
\begin{equation}
  \label{eq:34}
  [D, \pi(f)] \,\psi= -i\sqrt 2\left(\begin{array}{cc} 0
      & \ll(\bar\partial f) \\ \ll(\partial f) & 0\end{array} \right)
  \psi =-i\sqrt 2\left(\begin{array}{c} 
      \bar\partial f\star \psi_2\\ \partial f\star \psi_1\end{array} \right).
\end{equation}
Easy calculation \cite[eq. 3.7]{Cagnache:2009oe} yields
  \begin{equation}
    \label{eq:36}
    \norm{[D, \pi(f)]}= \sqrt 2\;\max \left\{ \norm{\ll(\partial f)},\, \norm{\ll(\bar\partial
        f)}\right\}.
  \end{equation}
There is no easy formula for the operator norm of $\ll$: unlike the commutative case,
$\norm{\ll(f)}$ is not the essential supremum of $f$. That is why (\ref{eq:36}) is not
very useful for explicit computations, and  one gets a more tractable formula using the Schr\"odinger representation. To this aim, and to make the link with familiar
notions of quantum mechanics, it is convenient to enlarge the algebra.

\subsection{Coordinate operators}
\label{coorsec} 
Obviously, the (unbounded) Moyal coordinate operators $\psi \to x_\mu\star
\psi$  do not belong to $\A$. So to correctly capture the geometry of
the Moyal plane, bigger algebras should be
considered, such as  the multiplier algebra $\M=\M_L\cap \M_R$ where
\begin{eqnarray}
\M_L=\{ T\in S ^{\prime} (\R^2)\;|\; T\star h\in S(\R^2)\; \text{for all}\;h\in S(\R^2)\},\\
\M_R=\{ T\in S ^{\prime} (\R^2)\;|\; h\star T\in S(\R^2)\; \text{for all}\;h\in S(\R^2)\}.
\end{eqnarray}
Here, the Moyal product is extended by
continuity to the dual  of $S(\R^2)$ as
$( T\star f,g)\doteq (T,f\star g)$ for $T\in \ss^{\prime}(\R^2)$ (and
analogously for $f\star T$ and the involution). 
$\M$ contains in particular \cite{Bondia:1988nr} 
the coordinate operators $x_\mu$.

A convenient representation of these algebras is
provided by Wigner
transition eigenfunctions,
\begin{equation}
  \label{eq:40}
   m,n\in\N, \qquad h_{mn} \doteq \frac 1{(\theta^{m+n}\, m!\, n!)^{\frac 12}} \bar z^{\star
    m}\star h_{00} \star  z^{\star
    n},\quad h_{00}= \sqrt{\frac{2}{\pi\theta}}e^{-\frac{(x_1^2 + x_2^2)}{\theta}}.
\end{equation}
They form an orthonormal basis of $L^2(\R^2)$ (see
\cite{Bondia:1988nr}, noticing that our $h_{mn}$ is their
$\frac{f_{mn}}{\sqrt{2\pi\theta}}$).
\begin{equation}
  \label{eq:43}
 h_{mn}\star h_{pq} = \frac{\delta_{np}}{\sqrt{2\pi\theta}}\, h_{mq},\quad h_{mn}^* =
  h_{nm},\quad (h_{mn}, h_{kl})= \delta_{mk} \delta_{nl}.
\end{equation}
The linear span $\dd$ of the $h_{mn}$'s for $m,n\in\N$
constitutes an
invariant dense domain of analytic vectors for the unbounded operators
$\ll(z), \ll(\bar{z})$, whose action writes 
\cite[Prop.~5]{Bondia:1988nr}
\begin{equation}
  \label{eq:44}
  \ll(z)\, h_{mn} = \sqrt{\theta m } \, h_{m-1,n},\quad   \ll(\bar z) \,
  h_{mn} = \sqrt{\theta(m+1)} \, h_{m+1,n}.
\end{equation} 
The same is true for the symmetric
operators $\ll(x_i)$, $i=1,2$.
By virtue of a
theorem of Nelson \cite{Reed1975}, these operators are essentially self-adjoint on $\dd$
($i.e.$ $\dd$ is a core for them all). Since $\dd\subset
S(\R^2)\subset L^2(\R^2)$, $S(\R^2)$ is as well a core for all of
them. On this domain, using 
$x_1 \star f=\left( x_1 f+i\frac{\theta}{2}\partial_{2}f\right)$ and
similar equations for $x_2\star f$ and $f\star x_i$ (see
\cite{Figueroa:2001fk})\cite{Bondia:1988nr}, one
obtains a representation of the Heisenberg algebra:
\begin{equation}
  \label{eq:70}
  [\ll(x_1),  \ll(x_2)]= i\theta\ii.
\end{equation}

\subsection{Schr\"odinger representation}

We make clear  the relation between the left-regular and the
Schr\"odinger representations, that is implicit in
\eqref{eq:70}. 
To make the dependence
on $\theta$ (identified to $\hbar$) explicit, we use the standard physicists
normalizations for the Schr\"odinger position and momentum operators, 
\begin{equation}
\frak q: (\frak q\psi)(x) = x\psi(x),\quad \frak p:
(\frak p\psi)(x)=-i\theta\partial_x\psi_{\lvert x},\quad \psi\in L^2(\R), x\in\R,
\label{eq:62}
\end{equation}
but we define the annihilation and creation operators as
\begin{equation}\label{eq:57}
\mathfrak a\doteq
\frac 1{\sqrt 2}(\frak q + i\frak p),\qquad\qquad \mathfrak a^*\doteq
\frac 1{\sqrt 2}(\frak q - i\frak p).
\end{equation}
This differs  from the usual convention, based on
dimensionless operators. In particular we have
    \begin{equation}
[\mathfrak a, \mathfrak a^*] = \theta \ii.
\label{eq:72}
\end{equation}
The eigenfunctions of the Hamiltonian
$\frak H\doteq \mathfrak a^*\mathfrak a+\theta/2\ii\;
\label{eq:88}$
 are then
\cite[$B_{V}$.(35) with $m=\omega=1$]{Cohen-Tannoudji:1973fk}
\begin{equation}
h_n(x)= (\theta\pi)^{-\frac 14} (2^n\,  n!)^{-\frac 12}\,
e^{-\frac{x^2}{2\theta}} H_n(\frac x{\sqrt{\theta}}),\quad n\in\N
\label{eq:77}
\end{equation}
where the $H_n$'s are the Hermite polynomials. The set
$\left\{h_n = \frac{(\frak a^*)^n}{\sqrt{\theta^n n!}}h_0\right\}$,
$n\in\N$, is an orthonormal basis of $L^2(\R)$
and spans an invariant dense domain $\dd_S$ of analytic
vectors for the operators $\frak q, \frak p$.

It is well known (see e.g. \cite[Theo. 2]{Bondia:1988qv}) that the
unitary operator $W: L^2(\R^2)\to L^2(\R)\otimes L^2(\R)$,
\begin{equation}
Wh_{mn}=h_m\otimes h_n \qquad m,n\in\N
\end{equation}
intertwines the left regular representation with the integrated Schr\"odinger representation
\begin{equation}
\label{repscro}
 \pi_S(f)\doteq \int \hat{f}(k_1,k_2)e^{\frac{i}{\theta}(\frak
   q k_1+\frak p k_2)}dk_1 dk_2.
\end{equation}
That is: $W\dd=\dd_S\otimes\dd_S$ and
$W\ll(f)W^*=\pi_S(f)\otimes\ii$
 for any $f\in S(\R^2)$.
In particular 
\begin{align}\label{W1}
 W&\ll(\bar{z})W^*= \mathfrak a^*\otimes\ii  &W&\ll(z)W^*= \mathfrak a\otimes\ii\\
W&\ll(x_1)W^*= \frak q\otimes\ii  &W&\ll(x_2)W^*= \frak p\otimes\ii . \label{WX}
\end{align}

The representation $\pi$ of the spectral
triple $T$ in (\ref{eq:23}) is a multiple of $\ll$, which in turn is unitary
equivalent{\footnote{Our normalization for $h_{mn}$,
    $h_m$ yields the Schr\"odinger representation without the
normalization term $\sqrt 2$ of~\cite{Bondia:1988nr}.}} 
to a multiple of the integrated Schr\"odinger
representation. Therefore,  for any $f\in\A$,
\begin{equation}
  \label{eq:9}
\norm{\ll(f)} = \norm{\pi(f)} = \norm{\pi_S(f)},
\end{equation}
 and we can denote the corresponding $C^*$-closure
 with the representation-free notation
\begin{equation}
  \label{eq:26}
  \bar\A \doteq \overline{\ll(\A)} \simeq \overline{\pi_S(\A)} \simeq \overline{\pi(\A)}. 
\end{equation}
As well known (see e.g. \cite{Bondia:1988qv}), this closure is isomorphic to the algebra of compact
operators,
\begin{equation}
\bar \A \simeq \kk.\label{eq:122}
\end{equation}

\subsection{Translations}
\label{transec}
We collect some
notations regarding translations, that is the transformation 
\begin{equation}
\label{defalpha}
(\alpha_{\kappa} f)(x) \doteq   f(x+\kappa)
\end{equation}
with $f\in S(\R^2)$ and $\kappa, x\in \R^2$.
Obviously $f_\kappa\doteq \alpha_{\kappa} f$ is Schwartz and
$f_\kappa\star g_\kappa(x) =
 (f\star g)_\kappa(x)$,
so that $\alpha_\kappa$ is a $*$-automorphism of the Moyal algebra $\A$.
In  the left-regular representation, it is implemented by the adjoint action of the plane wave with wave vector
    $\frac 1{\theta}\Theta\kappa$: one checks by easy calculation that for  $f\in S(\R^2), \kappa\in\R^2$ 
\begin{equation}
\label{defukappa}
\ll(\alpha_\kappa f) = \text{Ad}\, U_\kappa\,\ll(f) 
\quad \text{ with }\quad
  U_\kappa \doteq \ll(e^{\frac i{\theta} \cdot {\Theta\kappa}}).
\end{equation}
As operators on $S(\R^2)$, one has
\begin{equation}\label{derleftreg}
\ll(\kappa^\mu\partial_\mu f)=i\left[\ll\left(\frac{x\Theta\kappa}{\theta}\right), \ll(f)\right].
\end{equation}

{\rem
\label{tranlatecoord} $\text{Ad } \ll(e^{i\kappa \cdot})$ extends naturally to the
  multiplier algebra $\M$. In particular, by exponentiating
  \eqref{eq:70} we obtain as operators on $S(\R^2)$
 \begin{equation}\label{adj1}
\alpha_\kappa z = \text{Ad }\ll(e^{i\kappa \cdot})\,z =  z +
\frac{\kappa}{\sqrt 2},\qquad\qquad \alpha_\kappa \bar z = \text{Ad }
\ll(e^{i\kappa \cdot})\, \bar z = \bar z + \frac{\bar\kappa}{\sqrt 2}
\end{equation}
where $\bar\kappa$ is the complex conjugate of  $\kappa=(\kappa_1,
\kappa_2)\in\R^2$ identified  to $\kappa_1 +
i\kappa_2\in\C$. }
\newline

In the Schr\"odinger representation, (\ref{defukappa}) becomes
 \begin{equation}
    \label{eq:82}
    \pi_S(\alpha_\kappa f) = \text{Ad}\, \frak
    u_\kappa\;\pi_S(f)\quad \text{ where }\quad   \frak u_\kappa = e^{\frac { \bar\kappa \frak a -\kappa \frak a^* }{\theta\sqrt 2}}.
 \end{equation}
while (\ref{derleftreg}) yields
\begin{equation}
  \label{eq:100}
\quad \pi_S(\kappa^\mu\partial_\mu f)= [\frac{ \bar\kappa \frak
      a - \kappa \frak a^*}{\theta\sqrt 2}, \pi_S(f)].
\end{equation}

This yields the more tractable characterization of the Lipschitz
ball we were asking for below (\ref{eq:36}):
\begin{lem}
\label{lipschro}
A Schwartz function $f\in\A$ is in the Lipschitz ball $\bb_{\text{Lip}}(T)$ of the
spectral triple (\ref{eq:184}) of the Moyal plane if and only if
\begin{equation}
  \label{eq:49}
  \max \left\{ 
\norm{ [\frak{a}^*,\pi_S(f)] },
\norm{ [\frak{a},\pi_S(f)]} \right\}\leq \frac{\theta}{\sqrt 2}.
\end{equation}
\end{lem}
\begin{preuve}
From (\ref{eq:100}) with $\kappa =1,i$, one checks that
$\pi_S(\partial_x f) = \frac i{\theta}[\frak p,\pi_S(f)]$ and
$\pi_S(\partial_y f) = \frac {-i}{\theta}[\frak q,\pi_S(f)]$. Therefore
\begin{equation}
  \label{eq:135}
  \pi_S(\partial f) = \frac {-1}{\theta}[\frak a^*,\pi_S(f)], \quad
  \pi_S(\bar \partial f) = \frac {1}{\theta}[\frak a,\pi_S(f)].  
\end{equation}
The result follows from 
 (\ref{eq:36}) together with (\ref{eq:9}).
\end{preuve}

\section{Spectral distance between translated states}
\label{distancetranslationsection}
This section contains the first main result of the paper, namely theorem
\ref{theomain} where we show eq. (\ref{eq:171}) on the orbit
\begin{equation}
\ccc(\varphi)\doteq
\left\{\varphi_\kappa, \kappa\in\R^2\right\}
\label{eq:162}
\end{equation}
of any state $\varphi$ under the action of the translation group.
We first show that the Euclidean distance is an upper bound for the
spectral distance. Then we then exhibit a sequence of elements in $\A$ that
attains this upper bound, called the \emph{optimal element}. Finally,
we discuss the result in the light of the commutative case.

\begin{defi}
\label{deftrans}  Given any state $\varphi\in\sa$ and  $\kappa\in\R^2\simeq \C$, the $\kappa$-translated of
    $\varphi$ is the state
    \begin{equation}
    \varphi_\kappa~\doteq~\varphi~\circ~\alpha_\kappa\label{eq:104}
    \end{equation}
where the translation $\alpha_\kappa$ is defined in
(\ref{defalpha}). We call $\abs{\kappa} =\sqrt{\kappa_1^2 + \kappa_2^2}$
the translation amplitude.
\end{defi}
Since the Dirac operator commutes with translations, one immediately
gets\footnote{A unitarily implemented automorphism
of $\A$ commuting with $D$
is an isometry of $\sa$ for the spectral distance (see
e.g. \cite{Bellissard:2010fk}, \cite{Martinetti:2001fk}).} that  the
spectral distance on
any orbit $\ccc(\varphi)$ is
invariant by translation: $d_D(\varphi, \tilde\varphi) =  d_D(\varphi_\kappa, \tilde\varphi_\kappa).$ 
However this gives no information on 
$d_D(\varphi, \varphi_\kappa)$. 

\subsection{Upper bound}
\label{secup}
 
\begin{lem}
\label{lemupper}
  For any $\varphi\in\sa, f\in\bb_{\text{Lip}}(T)$ and $t\in [0,1]$, let us define
\begin{equation}
  \label{eq:13}
  F(t) \doteq \varphi_{t\kappa}(f) = \varphi(\alpha_{t\kappa}f),
\end{equation}
where $\kappa = (\kappa^1, \kappa^2)\in\R^2$ is fixed. Then
\begin{equation}
  \label{eq:38}
    \frac{dF}{dt}_{\lvert t} =\kappa^\mu\varphi_{t\kappa}(\partial_\mu f).
\end{equation}
\end{lem}
\begin{preuve}
For $f$ a Schwartz function, let us write
\begin{equation}
 \dot f = \frac d{dt}\alpha_{t\kappa} f = \kappa^\mu
 \alpha_{t\kappa}\partial_\mu f
\end{equation}
and, for any non-zero real number $h$, 
\begin{equation}\quad f_h\doteq \frac{\alpha_{(t+h)\kappa} f -\alpha_{t\kappa} f}h.
\label{eq:89}
\end{equation}
Notice that $\dot f$ and $f_h$ are in
$S(\R^2)$. From definition \ref{deftrans}, the result amounts to show that 
\begin{equation}
\lim_{h\to 0}\varphi(f_h) =\varphi(\dot f).
\label{eq:127}
\end{equation}

By linearity and continuity of $\varphi$,  one has
\begin{equation}
\label{eq:127bis}
 \abs{\varphi(f_h) - \varphi(\dot f)} \leq  
\norm{\varphi}\norm{\ll(f_h) -\ll(\dot f)} \leq \norm{f_h -\dot f}_{L^2(\R^2)}  
\end{equation}
where we used that the operator norm is smaller than the $L^2$ norm \cite[Lemma
2.12]{Gayral:2004rc}.  
 Observe that $f_h$ tends to $\dot f$
in the $S(\R^2)$ topology, meaning
that for every $\epsilon >0$ and integer $i>0$ we can
choose $\delta >0$ such that for $|h|<\delta$ one has, for instance,  $(1+|x|^i) \abs{f_h(x) -\dot f(x)} \leq \epsilon$, that is 
\begin{equation}
\abs{f_h(x) -\dot f(x)} \leq \frac{\epsilon}{(1+|x|^i)}.
\end{equation}
By the dominated convergence theorem, $f_h\to\dot f$ in the $L^2$-topology, so (\ref{eq:127bis}) implies
(\ref{eq:127}).\end{preuve}

\begin{prop}
\label{borne}
For any $\kappa\in\C$ and $\varphi\in\sa$,  $d_D(\varphi, \varphi_\kappa) \leq \abs{\kappa}.$
\end{prop}
\begin{preuve} 
Let us denote $\kappa^{a=1}=\frac 1{\sqrt 2}\kappa$,  $\kappa^{a=2}=\frac
1{\sqrt 2}\bar\kappa$, $\partial_{a=1}=\partial$, $\tilde\partial_{a=2} =\bar\partial$. Inverting ~(\ref{eq:18}) yields
\begin{equation}
  \label{eq:67}
  \kappa^\mu\, \varphi(\alpha_{t\kappa} \partial_\mu f)  = 
\frac 1{\sqrt 2} \left( \kappa\, \varphi(\alpha_{t\kappa} \partial f) +
  \bar\kappa \varphi (\alpha_{t\kappa} \bar\partial f) \right) =\kappa^a \varphi(\alpha_{t\kappa} \partial_a f).
\end{equation}
By Cauchy-Schwarz and the continuity of $\varphi$, at any $t$ one has
\begin{equation}
  \label{eq:17}
 \abs{\kappa^\mu
\varphi (\alpha_{t\kappa}  \partial_\mu f)}
\leq \abs{\kappa}\sqrt{\sum_a \abs{\varphi (\alpha_{t\kappa} \tilde\partial_a f)}^2}
\leq \abs{\kappa}\sqrt{\sum_a \norm{\ll(\tilde\partial_a f)}^2}.
\end{equation} 

For $f$ in the Lipschitz ball, 
(\ref{eq:36}) gives $\norm{\tilde\partial_a  f}\leq \frac
1{\sqrt 2}$ for $\mu=1,2$. Lemma \ref{lemupper} together with
(\ref{eq:17}) yields
\begin{equation}
  \label{eq:21}
  \abs{\frac{dF}{dt}_{\lvert_t}}\leq \abs{\kappa}
\end{equation}
for any $t$. Hence \begin{equation}
  \label{eq:14}
  \abs{\varphi_\kappa(f) -\varphi(f)} = \abs{F(1) - F(0)} \leq \int_0^1 \abs{\frac
    {dF}{dt}}_{\lvert_t} dt \leq \abs{\kappa}.
\end{equation}
\end{preuve}

\subsection{Optimal element \& regularization at infinity}
\label{optsection}
Inspired by the analogy \cite{Rieffel:1999ec,dAndrea:2009xr} in the commutative case between the spectral distance and the
Monge-Kantorovich, let us introduce the
following definition, which makes sense whatever algebra (commutative or not).

\begin{defi}
\label{defopt}
Given a spectral triple $T_1= (\A_1, \hh_1, D_1)$,  we call \emph{optimal element for a pair of
states $(\varphi, \tilde\varphi)$} an element of $\bb_{\text{Lip}}(T_1)$ that attains the
supremum in~(\ref{eq:22}) or,  in case the supremum is not attained, a sequence of elements $a_n\in
\bb_{\text{Lip}}(T_1)$ such that
\begin{equation}
\lim_{n\rightarrow +\infty} \abs{\varphi(a_n) - \tilde\varphi(a_n)} =
d_{D_1}(\varphi,\tilde\varphi).
\label{eq:27}
\end{equation}
\end{defi} 

As a first guess, we consider as an optimal element for a pair $(\varphi,\varphi_\kappa)$, $\varphi\in\sa, \kappa\in\C$,  the function 
\begin{equation}
f_0(x_1, x_2) \doteq \frac 1{\sqrt 2} (ze^{-i\Xi}  + \bar ze^{i\Xi})\label{eq:73}
\end{equation}
where $\Xi\doteq \text{Arg } \kappa$ and $z, \bar z$ are
defined in~(\ref{defcoorz}). Obviously $\ll(f_0)$ satisfies the
commutator norm condition (\ref{eq:49}) since
\begin{equation}
  \label{eq:29}
\norm{[\frak a,  \pi_S(f_0)]} = \frac 1{\sqrt2}\norm{[\frak
  a,\frak a^*]} = \frac {\theta}{\sqrt 2}
\end{equation}
together with a similar equation for $\norm{[\frak a^*,
  \pi_S(f_0)]}$. Furthermore, with  ${\bf 1}$ the constant function
$x\to 1$, one obtains
\begin{equation}
  \label{eq:85}
  \alpha_\kappa f_0 = f_0 + \abs{\kappa}{\bf 1}.
\end{equation}
Therefore, assuming $\varphi(z) < \infty$ (in the sense of remark
\ref{remstate} below, that is assuming all the $\psi_i$'s are in the domain of
$\frak a$, and the sum (\ref{eq:1}) is finite), and working
in the minimal unitization of $\A$ one gets, as expected,
\begin{equation}
  \label{eq:86}
  \abs{\varphi_\kappa(f_0) - \varphi(f_0)} =   \abs{\varphi(\alpha_\kappa f_0)
    - \varphi(f_0)} = \varphi(\abs{\kappa}{\bf 1}) = \abs{\kappa}.
\end{equation}

The point is that $f_0$ is not an optimal element, for it is not in
the Moyal algebra $\A$ but in its multiplier algebra
$\M$. So we need to
regularize it by finding a sequence $\left\{f_n\right\}, n\in\N$, in
$\bb_{\text{Lip}}(T)$ with $T$ the spectral triple (\ref{eq:2}), and which converges to
$f_0$ in a suitable topology. In the 
following proposition, we build from $f_0$ a net of element $f_\beta$ contained in the
Lipschitz ball. We extract from
it the required optimal element $\left\{f_n\right\}$ in the next subsection. 

\begin{prop}
\label{reg}
  Let $\kappa = \abs{\kappa
}e^{i\Xi}$ be a fixed translation. For $\beta\in\R^{*+}$, let us define
  \begin{equation}
 f_\beta \doteq \frac 1{\sqrt 2}(z_\beta + z_\beta^* )\quad \text{
   where } \quad z_\beta
 \doteq ze^{-i\Xi} \star e_\star^{-\frac{\beta}{\theta}\bar{z}\star z}.\label{eq:98}
\end{equation}
where $e_\star$ denotes the $\star$-exponential. Then there exists $\gamma>0$ such that $f_\beta \in
\bb_{\text{Lip}}(T)$ for any $\beta\leq \gamma$.
\end{prop}
\begin{preuve}
 First, let us check that $f_\beta$ is in $\A$.  As a formal power serie
of operators, one has
\begin{equation}
  \label{eq:130}
  W \ll(e_\star^{-\frac{\beta}{\theta} \bar z \star z})W^* = e^{-\frac{\beta}{\theta}\frak n}\otimes \ii,
\end{equation}
where $\frak n \doteq \frak a^*\frak a$ is the number operator. In the Schr\"odinger representation, $\frak n$ is
 a diagonal matrix with generic term $n\theta$. Therefore for any
 $\beta\in (0,\infty)$, the operator $e^{-\frac{\beta}{\theta}\frak n}$
is a matrix with fast decay coefficient, meaning that the r.h.s. of
(\ref{eq:130}) is in $\pi_S(\A)\otimes \ii$ and  $e_\star^{-\frac{\beta}{\theta} \bar z
  \star z}$ is in $\A$. The same is true for $f_\beta$ since $z$ is in
the multiplier algebra $\M$ of $\A$ (see also \cite{GGBNV}). 

From now on we put $\theta=1$ and assume that $\Xi = 0$. By Lemma \ref{lipschro}, $f_\beta$ is in the Lipschitz ball if and only if
$\norm{\mathfrak{c}(\beta)}\leq 1\label{eq:215}$
where we define
\begin{equation}
\label{deff}
 \frak c(\beta) \doteq [\frak{a},\mathfrak{a}_\beta
 +\mathfrak{a}_\beta^*]= \left( e^{-\beta}-e^{\beta}(1-e^{-\beta})\frak{a}^2-(1-e^{-\beta})\frak{n}\right) e^{-\beta\frak{n}},
\end{equation}
with $\frak a_\beta \doteq \pi_S(z_\beta) = \mathfrak{a}
e^{-\beta \mathfrak{n}}$. The r.h.s. of (\ref{deff})
follows from the Baker-Campbell-Hausdorff formula,
$e^{-\beta \frak n} \frak a e^{\beta  \frak n} = \frak a
e^{\beta}$, that is $[\frak a, e^{-\beta \frak n}]= (1- e^\beta)  \frak
a e^{-\beta \frak n}$. In the energy eigenvectors basis, one gets
\begin{equation}
  \label{eq:230}
  \frak c(\beta)=\lp \begin{array}{ccccccccc}
\lambda_{0,0} &0&\lambda_{0,2} &0&0 &\dots&\dots&\dots\\
0& \lambda_{1,1}&0&\lambda_{1,3}&0&\ddots&\ddots&\ddots\\
\vdots&\ddots&\ddots&\ddots&\ddots&\ddots&\ddots&\ddots\\
\vdots&\ddots&\ddots&\ddots&0&\lambda_{n-2,n}&0&\ddots\\
\vdots&\ddots&\ddots&0&\lambda_{n-1,n-1}&0&\lambda_{n-1,n+1}&0\\
\vdots&\ddots&\ddots& \ddots& 0&\lambda_{n,n}&0&\lambda_{n,n+2}\\
\vdots&\ddots&\ddots&\ddots&\ddots&\ddots&\ddots&\ddots\\
\end{array}\rp
\end{equation}
where
\begin{eqnarray}
\lambda_{n,n} &=&\left(e^{-\beta}-(1-e^{-\beta})n\right)e^{-\beta n},\\
\lambda_{n-2,n} &=&-e^{\beta}(1-e^{-\beta})\sqrt{ n(n-1)}e^{-\beta n}.
\end{eqnarray}
 To estimate the norm of $\frak c(\beta)$ we use Schur's test, that is
\begin{equation}
\norm{\frak c(\beta}\leq\left(\underset{n}{\sup} \;\underset{m}{\sum}
  \abs{\lambda_{m,n}}\right)^{\frac 12} \left(\underset{m}{\sup} \;\underset{n}{\sum}
  \abs{\lambda_{m,n}}\right)^{\frac 12}.
\end{equation}
Actually we prove that  $\| \frak c(\beta)\|\leq \left(
   e^{-\beta}e^{\beta}\right)^{\frac{1}{2}}=1$ by showing that for $\beta$ sufficiently small
\begin{eqnarray}
&|\lambda_{n-2,n}|+|\lambda_{n,n}|\leq e^{-\beta },\label{dis1}\\
&\!\!\!|\lambda_{n,n+2}|+|\lambda_{n,n}|\leq e^{\beta}.\label{dis2}
\end{eqnarray}

 Let us begin with (\ref{dis1}). For $n\leq
 e^{-\beta}/(1-e^{-\beta})\doteq n_0 \in \R$ one has $\lambda_{n,n}\geq0$,
 while $\lambda_{n-2,n}\leq0$ for any
 $n\in\N$. So
\begin{equation*}
|\lambda_{n-2,n}|+|\lambda_{n,n}|=
\begin{cases}
\left( e^{-\beta}+(1-e^{-\beta})(e^{\beta}\sqrt{n(n-1)}-n)\right)e^{-\beta n} & \text{for}\quad n\leq n_0,\\
\left( -e^{-\beta}+(1-e^{-\beta})(e^{\beta}\sqrt{n(n-1)}+n)\right)e^{-\beta n} & \text{for}\quad n> n_0.
\end{cases}
\end{equation*}

Let us assume $n\leq n_0$. Then
$e^{\beta}\sqrt{n(n-1)}-n\leq0$ for $n\leq (1-e^{-2\beta })^{-1}
\doteq n_1 \in\R$. Since $n_1 \leq n_0$ as soon as
\begin{equation}
\beta \leq \beta_0\doteq
\ln{\left((1+\sqrt{5})/2\right)},\label{eq:142}
\end{equation}
\eqref{dis1} is true for  $\beta\leq\beta_0$ and $0\leq n\leq n_1$.
For $\beta\leq\beta_0$ and $n_1\leq n\leq n_0$, we have
\begin{equation*}
|\lambda_{n-2,n}|+|\lambda_{n,n}|\leq \left( e^{-\beta}+n(1-e^{-\beta})(e^{\beta}-1)\right)e^{-\beta n}\leq \left( e^{-\beta}+e^{-\beta}(e^{\beta}-1)\right) e^{-\beta/(1-e^{-2\beta})}\leq e^{-\beta},
\end{equation*}
where we simply substitute  for $n_0$ in the polynomial factor and for $n_1$ in the exponential.

Suppose now $n > n_0$, so that
\begin{equation}
|\lambda_{n-2,n}|+|\lambda_{n,n}|\leq \left( -e^{-\beta}+ n(1-e^{-\beta})(e^{\beta}+1)\right)e^{-\beta n}.
\end{equation}
The function $(an-b)e^{-\beta n}$ ($a,b>0$) reaches its maximum for
$n=\beta^{-1}+b/a$, therefore 
\begin{align*}
\nonumber
|\lambda_{n-2,n}|+|\lambda_{n,n}|&\leq\left( -e^{-\beta}+(1-e^{-\beta})(e^{\beta}+1)\left(\frac{1}{\beta}+\frac{e^{-\beta}}{(1-e^{-\beta})(e^{\beta}+1)}\right)\right)e^{-\beta \left(\frac{1}{\beta}+\frac{e^{-\beta}}{(1-e^{-\beta})(e^{\beta}+1)}\right)}\\
&=\frac{(1-e^{-\beta})(e^{\beta}+1)}{\beta}\, e^{-1}\, e^{
  -\frac{\beta e^{-\beta}}{(1-e^{-\beta})(e^{\beta}+1)}} \leq (e^{\beta}+1)e^{-1}
\end{align*}
where we use $1-e^{-\beta}\leq \beta$ for any $\beta \in\R^+$. One
checks that $e^{\beta}(e^{\beta}+1)\leq e$ as soon as
\begin{equation}
\beta\leq\beta_1\doteq \ln\lp \frac 12(\sqrt{1+
  4e}-1\rp.\label{eq:190}
\end{equation}
Consequently, whatever $n$ the inequality \eqref{dis1} is true for  $\beta\leq
\text{min}\lp\beta_0,\, \beta_1\rp = \beta_1$.
\newline

We now show \eqref{dis2}. For any $\beta, n:$ $\lambda_{n,n+2}=
-e^{-\beta}(1-e^{-\beta})\sqrt{(n+1)(n+2)}  e^{-\beta n}\leq 0$. So
\begin{equation*}
|\lambda_{n,n+2}|+|\lambda_{n,n}|=
\begin{cases}
\left( e^{-\beta}+(1-e^{-\beta})(e^{-\beta}\sqrt{(n+1)(n+2)}-n)\right)e^{-\beta n} & \text{for}\quad n\leq n_0,\\
\left( -e^{-\beta}+(1-e^{-\beta})(e^{-\beta}\sqrt{(n+1)(n+2)}+n)\right)e^{-\beta n} & \text{for}\quad n> n_0.
\end{cases}
\end{equation*}

For $n=0$, this yields 
\begin{equation}
  \label{eq:193}
  |\lambda_{0,2}|+|\lambda_{0,0}| = e^{-\beta} +
  (1-e^{-\beta})e^{-\beta}\sqrt{2}\leq e^{-\beta} \lp 1+ \beta \sqrt{2}\rp,
\end{equation}
 which is obviously smaller than $e^{\beta}$ since
 $1+\sqrt{2}\beta\leq e^{2\beta}$ for any $\beta\in\R^+$.

For $1\leq n\leq n_0$, either $(e^{-\beta}\sqrt{(n+1)(n+2)}-n)\leq 0$
and we are done; or
\begin{equation}
\label{casdeux}
\lp e^{-\beta}+(1-e^{-\beta})(e^{-\beta}\sqrt{(n+1)(n+2)}-n) \rp e^{-\beta n}\leq
\lp e^{-\beta}+\beta (\sqrt{(n+1)(n+2)}-n)\rp e^{-\beta}.
\end{equation}
Observing that
\begin{equation}
\sqrt{(n+1)(n+2)}-n= \frac{3n+2}{\sqrt{(n+1)(n+2)}+n}\leq \frac{3n+2}{2n+1}
\leq \frac{3}{2}+\frac{1}{n}
\leq \frac{5}{2},
\end{equation}
one obtains that (\ref{casdeux}) is smaller than $e^{\beta}$ as soon as
$e^{-\beta}+5\beta/2\leq  e^{2\beta}$. Noticing that $e^{2\beta}\geq 1
+2\beta$, this is true as soon as 
$e^{-\beta}+5\beta/2\leq  1+2\beta$, or equivalently as soon as $(\beta -2)e^{\beta-2} \leq -2e^{-2}$, that is - denoting $W$ the
Lambert function - for
\begin{equation}
  \label{eq:192}
  \beta\leq \beta_2 = 2+ W(-2e^{-2}).
\end{equation}

We are left with the case $n\geq n_0$. Then
\begin{align}
 |\lambda_{n,n+2}|+|\lambda_{n,n}| 
&\leq \left( -e^{-\beta}+(1-e^{-\beta})
  (e^{-\beta}(n+2)+n)\right)e^{-\beta n}\\ &=\left(
  (1-e^{-2\beta})n  - (2e^{-2\beta} -e^{-\beta})\right)e^{-\beta n}. 
\end{align}
By the same reasoning as before, this is maximum for $n=\beta^{-1}+
(2e^{-2\beta}-e^{-\beta})/(1-e^{-2\beta})$, so that 
\begin{equation}
\label{melenchon}
  |\lambda_{n,n+2}|+|\lambda_{n,n}| \leq \frac{1-e^{-2\beta}}{\beta}
  e^{-1}e^{-\beta\frac{2e^{-2\beta}-e^{-\beta}}{1-e^{-2\beta}}} \leq
  \frac{2\beta}{\beta}e^{-1} =\frac 2e<e^{\beta} \quad \forall
  \beta\in \R^{+}.
\end{equation}

To summarize,  $\norm{c(\beta)}\leq 1$  for any $n\in\N$ as soon as
$\beta \leq \gamma\doteq  \text{min} (\beta_1, \beta_2) =\beta_1.$ 
To conclude, we notice that restoring $\theta$ and $\Xi$ simply amounts to multiply $\lambda_{n,n}$ by $\theta e^{-\Xi}$, and
$\lambda_{n,n+2}$ by   $\theta e^{-\Xi}$, letting the proof unchanged.
\end{preuve}

\subsection{Main result}
\label{subsecmain}

At this point it might be useful to recall some well known facts regarding the state space of $\bar \A$. By (\ref{eq:122}) and  a classical result
of operator algebras (see for example \cite{Sakai1971}),  in every representation of $\bar \A$ all states are normal, while all pure states
are actually vector states. When the representation is irreducible (like the integrated Schr\"odinger representation), the
correspondence between pure and vector states becomes one to one. In addition, normality has the following important consequence.
\begin{rem}
1. \label{remstate} 
  Any non-pure state $\varphi\in \sa$ is a convex
 combination of pure states,
 \begin{equation}
\varphi(a) = \sum_{n=1}^{\infty}
 \lambda_i \, \scl{\psi_i}{\pi_S(a)\psi_i}\quad \quad \forall a\in \A,
\label{eq:1}
 \end{equation}
where $\psi_i$ are unit vectors in $L^2(\R)$ and $\lambda_i\in\R^+$ with $\sum_{i=1}^{\infty}
 \lambda_i =1$ \cite[Theo. 7.1.12]{Kadison1986}. Consequently, the restriction of $\varphi$  to the closed ball of
  radius $r\in \R^{*+}$,
 $\bb_r(\A)\doteq \left\{a\in\A, \norm{a}\leq r\right\},$
\label{eq:113}
can be approximated by a \textit{finite} combination of pure states. Indeed, denoting $n_\epsilon$ the smallest integer
 such that $\sum_{i=n_\epsilon+1}^\infty \lambda_i \leq \epsilon$ for some arbitrary fixed $\epsilon$, one has
\begin{equation}
  \label{eq:109}
\abs{\varphi(a)-\sum_{n=1}^{n_\epsilon}
 \lambda_i \scl{\psi_i}{\pi_S(a)\psi_i}}\leq r\epsilon \quad \quad \forall a\in \bb_r(\A).
\end{equation} 
2. We denote ${\mathcal S} _0(\A)\subset\sa$ the convex hull of the Schwartz pure
states of $\A$, that is $\varphi=\scl{\psi}{\cdot\psi}$, $\psi\in {\cal S}(\R)$. 
\end{rem}

We can now prove the first main result of this paper, namely eq.(\ref{eq:171}) for
any state $\varphi$ in $\sa$ and any translation $\kappa\in\R^2\simeq\C$.

\begin{thm}
\label{theomain}
  The spectral distance between a state and its translated is the Euclidean distance,
  \begin{equation}
 d_D(\varphi, \varphi_\kappa) = \abs{\kappa} \quad\quad\forall\; \varphi\in\ss(\A),\;
 \kappa\in\C.
\label{eq:71}
  \end{equation}
\end{thm}
\begin{preuve}
We split the proof in three parts: first we show that
the result follows if
\begin{equation}
\lim_{n\to \infty} \varphi(A(\beta_n)_{t\kappa})= 0
\label{eq:56}
\end{equation}
where $\left\{\beta_n\right\}$ is a sequence of positive numbers
tending to $0$, satisfying
$\beta_n\leq \gamma \;\forall n\in\N$ for $\gamma$ introduced in proposition \ref{reg}.  $A(\beta)_{t\kappa}$ 
is defined below. Then we show that
(\ref{eq:56}) actually holds for pure states. Finally we extend the result to arbitrary states.
\newline

i) Let us fix $\beta>0$ and consider the net $f_\beta$,
$0<\beta\leq\gamma$  in the Lipschitz ball
defined in (\ref{eq:98}). To lighten notation, we incorporate $\theta$
into $\beta$, i.e. $\frac{\beta}{\theta}\to\beta$. The theorem amounts to show that, for any any state
$\varphi\in\sa$ and any $\kappa\in\C$, one has
\begin{equation}
  \label{eq:16}
  \lim_{\beta\to 0}\; \abs{\varphi_\kappa(f_\beta) - \varphi(f_\beta)}= \abs{\kappa}.
\end{equation}
Defining
$F(t) \doteq \varphi_{t\kappa}(f_\beta) =
\varphi(\alpha_{t\kappa}f_\beta),
\label{eq:66}$
we will be done as soon as we show that {\footnote{Notice that the pointwise limit is sufficient: substituting in (\ref{eq:16})
$\abs{\varphi_\kappa(f_\beta) - \varphi(f_\beta)}$ with its integral
form (\ref{eq:14}), $f_\beta$ being in
the Lipschitz ball allows to exchange the limit and the integral
thanks to the dominated convergence theorem.}}
\begin{equation}
  \label{eq:64}
 \lim_{\beta\to 0} \,\frac{dF}{dt} = \abs{\kappa}.
\end{equation}

To this aim, let us fix $\kappa\in\C$. By lemma \ref{lemupper} one has
\begin{equation}
\label{dfgrossen}
\frac{d F}{dt}_{\lvert_t}= \kappa^\mu\varphi_{t\kappa}(\partial_\mu f_\beta)
=\kappa^\mu\varphi(\partial_\mu (\alpha_{t\kappa}f_\beta)).
\end{equation}
Using that $\bar \kappa \frak{a}-\kappa\frak{a}^*$ commutes
with $\frak u_\kappa$, (\ref{eq:82}) yields
\begin{equation}
\label{house}
\kappa^\mu\pi_S(\partial_\mu(\alpha_{t\kappa}f_\beta)) =
\left[\frac {\bar \kappa \frak{a}-\kappa\frak{a}^*}{\theta\sqrt 2},\pi_S(\alpha_{t\kappa}f_\beta)\right] =
\text{Ad}\, \frak u_{t\kappa}\left[\frac {\bar \kappa \frak{a}-\kappa\frak{a}^*}{\theta\sqrt 2} ,\pi_S(f_\beta)\right].
\end{equation}
Thus, denoting $\frak f_\beta \doteq
\pi_S (f_\beta)$, 
\begin{equation}
\label{grossen2}
\frac{d F}{dt}_{\lvert_t}= \frac 1{\theta\sqrt 2}\;\varphi(\text{Ad}\,
\frak u_{t\kappa}\left[ \bar \kappa \frak{a}-\kappa\frak{a}^*\,,\, \frak f_\beta\right]),
\end{equation}
where, with a slight abuse of notation, we write the evaluation of a
state as  $\varphi(\pi_S(f))$
instead of $\varphi(f)$. 
By easy computations, one has
\begin{align}
\left[\bar \kappa \frak{a}-\kappa\frak{a}^*\,,\, \frak f_\beta\right] &=
\frac 1{\sqrt 2}(\left[\bar \kappa \frak{a}\,,\, \frak a_\beta \right]+ \left[\bar
  \kappa \frak{a}\,,\, \frak a_\beta^*  \right])
+ \text{adjoint},\\
& \label{grossen}
=\frac 1 {\sqrt 2}\left( \theta\abs{\kappa}e^{-\beta \frak n} + \bar \kappa e^{-i\Xi}\frak a\left[ \frak{a}\,,\,e^{-\beta \frak n}\right] + \bar\kappa e^{i\Xi}\left[
  \frak{a}\,,\,e^{-\beta\frak n}\right]\frak a^* +
\text{adjoint}\right).\end{align}
Let us denote the sum of the commutators in the equation above as a
single operator $A(\beta)$, which 
is in $\A$ since 
both $e^{-\beta \frak n}$ and  $\left[\bar \kappa
  \frak{a}-\kappa\frak{a}^*\,,\, \frak f_\beta\right] = \theta\sqrt 2
\kappa^\mu\partial_\mu \frak f_\beta$ are in $\A$. 
Define similarly
\begin{equation}
  \label{eq:107}
  A(\beta)_{t\kappa} \doteq \text{Ad}\, \frak u_{t\kappa} \; A(\beta) =
  \text{Ad}\, \frak u_{t\kappa}\, \left[ \bar \kappa \frak{a}-\kappa\frak{a}^*\,,\,
    \frak f_\beta\right] - \sqrt 2\theta\abs{\kappa}e^{-\beta \frak n _{t\kappa}},
\end{equation}
with
\begin{equation}
  \label{eq:106}
 \frak a_{t\kappa}\doteq  (\text{Ad}\,\frak u_{t\kappa})\frak a = \frak
 a + \frac{t\kappa}{\sqrt 2}\ii, \quad \frak a^*_{t\kappa}\doteq
 (\text{Ad}\,\frak u_{t\kappa})\frak a^*= \frak a^*+ \frac{\bar t\kappa}{\sqrt 2}\ii,\quad
\frak n _{t\kappa}\doteq (\frak a^* \frak a)_{t\kappa} = \frak a^*_{t\kappa}\frak a_{t\kappa}.
\end{equation} 
Again $A(\beta)_{t\kappa}$ is in $\A$, for the latter is invariant by
$\text{ ad}\, \frak
u_{t\kappa}$. 
This allows to write (\ref{grossen2}) as
\begin{equation}
   \label{eq:108}
  \frac{d F}{dt}_{\lvert_t}= |\kappa|\,\varphi(e^{-\beta\frak
    n_{t\kappa}}) + \frac 1{\theta\sqrt 2}\varphi(A(\beta)_{t\kappa}).
 \end{equation}

The operator
$\frak n_{t\kappa}$ is positive and selfadjoint, so  by the Hille-Yosida theorem
\cite{Reed1975} the application
$(0,+\infty)\ni
\beta\to e^{-\beta n_{t\kappa}}$
defines a contraction semi-group.
In particular  one has 
for $\beta\geq 0$ and any $\psi\in L^2(\R)$,
\begin{equation}
\norm{e^{-\beta n_{t\kappa}}}\leq 1 \quad \text{and }\quad
\label{strongcont}
\lim_{\beta\to 0} e^{-\beta \frak n _{t\kappa}}\psi=\psi,
\end{equation}
so that remark \ref{remstate} yields
\begin{equation}
\lim_{\beta\to 0} \varphi(e^{-\beta \frak n_{t\kappa}})= 1.\label{eq:102}
\end{equation}
As soon as the limit (\ref{eq:56}) holds
true for some sequence $0<\beta_n\leq \gamma$, (\ref{eq:108}) reduces to (\ref{eq:64}) and the theorem
follows.
\newline

ii) To prove the limit (\ref{eq:56}), we need to evaluate the various terms
of $\varphi(A(\beta)_{t\kappa})$. Let us first do it assuming
$\varphi$ is a pure state in $\s0a$. Developing
the commutator in (\ref{eq:107}), one obtains
\begin{equation}
\label{premierterme}
A(\beta)_{t\kappa}  = \frac 1{\sqrt 2}
\left(
\bar \kappa e^{-i\Xi}\frak a_{t\kappa}
\left[\frak a\,,\,e^{-\beta\frak n_{t\kappa}}\right] + 
\bar \kappa e^{i\Xi}
\left[\frak{a}\,,\,e^{-\beta \frak n_{t\kappa}} \right] \frak a_{t\kappa}^*
\right) +\text{adjoint.}
\end{equation}

Let us consider the first term of this equation, disregarding the constant
coefficients.  One has 
\begin{align}
\|\frak a_{t\kappa}\left[ \frak{a}\,,\,e^{-\beta\frak n_{t\kappa}}\right]\psi\| &= \|\frak a_{t\kappa}\left[ \frak a
+\frac{t\kappa}{\sqrt 2}\ii , \,\ii - e^{-\beta \frak n_{t\kappa}}\right]\psi\|\\
&
\label{secpnd}\leq \|\frak a _{t\kappa}^2( \ii - e^{-\beta\frak
  n_{t\kappa}})\psi\|+\|\frak a _{t\kappa} ( \ii - e^{-\beta \frak n_{t\kappa}}) \frak a _{t\kappa}\psi\|.
\end{align}
Calculating explicitly the first norm in (\ref{secpnd}), one finds
\begin{align}
\|\frak a^2 _{t\kappa}&( \ii - e^{-\beta\frak n_{t\kappa}})\psi\|^2 
= \langle \frak a^2_{t\kappa} ( \ii - e^{-\beta\frak n_{t\kappa}})\psi, \frak a^2 _{t\kappa} ( \ii - e^{-\beta \frak n_{t\kappa}})\psi\rangle \\
&=\langle \frak a^2 _{t\kappa} e^{-\beta \frak n_{t\kappa}}\psi, \frak
a^2 _{t\kappa} e^{-\beta \frak n_{t\kappa}}\psi\rangle + \langle \frak a
^2_{t\kappa}\psi, \frak a^2 _{t\kappa} \psi\rangle -2\text{Re} \langle \frak a^2 _{t\kappa} e^{-\beta \frak
  n_{t\kappa}}\psi, \frak a^2 _{t\kappa} \psi\rangle\\
\label{25octobre1}&= \langle e^{-2\beta \frak n_{t\kappa}}\psi, \frak a _{t\kappa}^{*2}
\frak a ^2_{t\kappa}\psi\rangle + \langle\psi,  \frak a_{t\kappa} ^{*2}\frak a^2 _{t\kappa} \psi\rangle 
-2\text{Re} \langle e^{-\beta \frak n_{t\kappa}}\psi,\frak a_{t\kappa}^{*2}\frak a^2_{t\kappa}\psi\rangle.
\end{align}
The three terms in (\ref{25octobre1}) are finite, for $\psi$ is
Schwartz. Moreover, by (\ref{strongcont}) they cancel each other as $\beta \to 0$. The same argument applies to $\|\frak a_{t\kappa}\left[ \frak{a}\,,\,e^{-\beta\frak
    n_{t\kappa}}\right] \frak a_{t\kappa}\psi\|$. Repeating the procedure
for $\left[\frak{a}\,,\,e^{-\beta \frak n_{t\kappa}} \right] \frak
a_{t\kappa}^*$ and the adjoints, one gets
\begin{equation}
  \label{eq:137}
  \lim_{\beta\to 0} \norm{A(\beta)_{t\kappa}\psi}=0,
\end{equation} 
so that, by
Cauchy-Schwarz,
$\lim_{\beta\to 0} |\varphi(A(\beta)_{t\kappa})|\leq \lim_{\beta\to 0} \norm{A(\beta)_{t\kappa}\psi}=0.$
This implies (\ref{eq:56}) and the result.
\newline

Now, fix any pure state $\tilde\varphi=
\scl{\tilde\psi}{\cdot \tilde\psi}$ for some unit vector
$\tilde\psi\in L^2(\R)$, and take a Schwartz-pure state $\varphi$ as before
such that
\begin{equation}
\label{limstate}
\|\varphi-\tilde\varphi\|<\frac{\epsilon}{r}
\end{equation}
for
arbitrary real positive numbers $r$ and $\epsilon$. 
 This is always possible for $S(\R)$ is dense
in $L^2(\R)$ (by Cauchy-Schwarz one has $\abs{(\varphi-\tilde\varphi)(a)}\leq 2\norm{\psi}_{L^2(\R)}\norm{\delta\psi}_{L^2(\R)} +
  \norm{\delta \psi}_{L^2(\R)}^2$ for any $a$
of norm $1$,  where $\delta\psi \doteq \tilde\psi -\psi$ has arbitrary
small norm). Then
\begin{equation}
\label{fin}
\abs{\tilde\varphi(A(\beta)_{t\kappa})}\leq \norm{\tilde\varphi-\varphi}\norm{A(\beta) _{t\kappa}} +
\abs{\varphi(A(\beta) _{t\kappa})}\leq \frac{\epsilon}r
  \norm{A(\beta) _{t\kappa}} + \abs{\varphi(A(\beta) _{t\kappa})}.
\end{equation}

From the definition (\ref{eq:107}) of $A(\beta)_{t\kappa}$,
the explicit form (\ref{house}) of the derivative  and the strong
continuity (\ref{strongcont}), using  moreover that $f_\beta$ is in the
Lipschitz ball so that - by (\ref{eq:36}) -  $\norm{\partial_\mu \frak
  f_\beta}\leq 2^{-\frac 12}$, one obtains\begin{equation}
\norm{A(\beta)_{t\kappa}} \leq \theta\sqrt 2 \kappa^\mu\norm{\partial_\mu f_\beta} +  \sqrt 2 \theta\abs{\kappa}
\leq \theta\sum_\mu \abs{\kappa^\mu}  +
\sqrt 2 \theta|\kappa|.
\end{equation}
Taking as a parameter $r$ the r.h.s. of
the equation above, 
\begin{equation}
  \label{eq:138}
  r = \theta\sum_\mu \abs{\kappa^\mu}  +
\sqrt 2 \theta|\kappa|,
\end{equation}
and remembering, as shown above,  that eq.(\ref{eq:56}) holds
true for $\varphi$, eq.(\ref{fin}) yields
\begin{equation*}
 \lim_{\beta\to
   0}\,\abs{\tilde\varphi(A(\beta)_{t\kappa})}=0,
\end{equation*}
hence the result.
\newline

iii) The argument for an arbitrary state in $\sa$ is now
straightforward.
For any $t\in[0,1]$, the net
$A(\beta)_{t\kappa}$, $0<\beta\leq \gamma$, is contained within the
closed ball $\bb_r(\A)\subset \bb(L^2(\R))$ with radius $r$ given by
(\ref{eq:138}).
Since $\bb_r(\A)$ is compact (and metrizable)  in the
$\sigma$-weak topology of $\bb(L^2(\R))$ (as any closed ball, see \cite{Sakai1971}),
from any sequence $\left\{ A(\beta_n)_{t\kappa} \right\}_{n=1}^{+\infty}$
with  $\beta_n\to 0$, one can extract a sub-sequence $\{A(\beta_{n_j})\}_{j=1}^{+\infty}$ such that,
for every (normal) state $\varphi$ in the predual $\bb(\hh)_*$,
\begin{equation}
\lim_{j\to \infty} \varphi(A(\beta_{n_j})) = \varphi(A(0))
\label{eq:101}
\end{equation}
for some $A(0)\in\bb_r(\A)$. Fixing $\epsilon>0$, the same is true for
the finite convex combination 
\begin{equation}
  \label{eq:139}
  \sigma_\epsilon \doteq \sum_{n=1}^{n_\epsilon}
 \lambda_i \scl{\psi_i}{\cdot\psi_i}
\end{equation} 
defined in remark \ref{remstate}, that is
\begin{equation}
\lim_{j\to \infty} \sigma_\epsilon(A(\beta_{n_j})) = \sigma_\epsilon
(A(0)).
\label{eq:20}
\end{equation}
But by  the result of ii), each of the terms of
$\sigma_\epsilon(A(\beta_{n_j}))$ tends to zero, that is 
$\sigma_\epsilon(A(0)) = 0$.
Therefore
\begin{equation}
  \label{eq:68}
  \abs{\varphi(A_0)}\leq r\epsilon,
\end{equation}
that is \begin{equation*}
\lim_{j\to \infty} \, \abs{\varphi(A(\beta_{n_j})}\leq r\epsilon 
\end{equation*}
which again is (\ref{eq:56}).
\end{preuve}

\subsection{Discussion on the optimal element}
\label{remradial}

In this paragraph, optimal always means ``up to
regularization''. The optimal element $f_0$~(\ref{eq:73})
is similar to the one used in \cite[Prop. 3.2]{dAndrea:2009xr} to compute the
distance between $\kappa$-translated states of $C_0(\R^2)$, $\kappa\in\R^2$, namely
\begin{equation}
  \label{eq:112}
  u\in\R^2 \to u \cdot \frac{\kappa}{\abs{\kappa}}
\end{equation}
with ``$\cdot$'' the scalar product of
$\R^2$. To get convinced, remember that the point $u=(u_1, u_2)\in\R^2$
identifies to
$z= \frac 1{\sqrt 2}( u_1 +i u_2)\in\C$ while the translation 
amplitude $\kappa = \abs{\kappa}(\cos\Xi,
\sin\Xi)$ identifies to $\abs{\kappa}e^{i\Xi}$ (see
(\ref{defcoorz}) and 
remark \ref{tranlatecoord}). With these conventions the r.h.s. of (\ref{eq:112}) is precisely (\ref{eq:73}):
\begin{equation}
  u_1\cos\Xi + u_2\sin\Xi = \sqrt 2\, \Re(ze^{-i\Xi}) = \frac 1{\sqrt
    2}(z e^{-i\Xi} + \bar z e^{i\Xi}).\label{eq:203}
\end{equation}
In other terms, an optimal element between translated states in the Moyal
plane is obtained as the left regular representation of an optimal element between
translated states in the commutative case. In the language of optimal
transport \cite{Villani:2003fk}, this means that
the optimal transportation plan between translated distributions does
not see the quantum nature of space. 
This is a striking result, that was not granted from
the beginning, in particular because not all the optimal elements of the commutative case give
an optimal element in the Moyal plane. 

Indeed, besides ~(\ref{eq:112}) another optimal element in the Euclidean
plane between a pure state $\delta_x$
and its translated $\delta_{y\doteq\lambda x}$, $\lambda\in\R^+$, is
the radial function
$\sqrt2\abs{z}$.
One could have thought there exists as well a ``radial'' optimal
element    (i.e. depending on $z\star \bar z$ only) between
translated states in the Moyal plane. It turns out that
  the supremum on such elements yields a quantity lower than the spectral distance  \cite{lizzif}. 

In proposition \ref{reg}, we regularize
by $e^{-\frac{\beta}{\theta}\abs{\ll(z)}^2}$,
 instead of $e^{-\frac{\sqrt{2}}n \abs{z}}$ in \cite{dAndrea:2009xr}. The
 constant coefficients in the exponential are not relevant, only the
 exponents of the argument of the exponential are. In the commutative case,
 one could choose as well $\abs{z}^2$ or
 $\abs{z}^3$,  for  $\norm{\text{grad}\lp\abs{z}e^{-\frac{\abs{z}^p}n}\rp}$ is never greater than $1$ for $p=1,2,3$. In proposition \ref{reg} we
chose the exponential $p=2$ rather than $p=1$, for the commutation
relations of $\abs{\ll(z)}^2 =\frak n$
are easier to deal with than those of $\abs{\ll(z)} =\sqrt \frak n
=\abs{\frak a}$.

\section{Symplectic orbits}
\label{symplectic}

One may wonder whether our methods apply beyond translations,
for instance to the orbit{\footnote{This notation generalizes the one used so far: for $M(t)$ a
    translation $t\kappa$, $\varphi_t$ in (\ref{eq:214})
    is $\varphi_{t\kappa}$ of definition \ref{deftrans}.}}
\begin{equation}
\varphi_t(f) \doteq \varphi\circ\alpha_{M(t)}(f) \quad\text{ with }\quad (\alpha_{M(t)}f)(x) = f(M(t)x)
\label{eq:214}
\end{equation}
of a state $\varphi$ under the action of a $1$-parameter subgroup
\begin{equation}
M(t)=e^{tJ}, \quad J=(S,\kappa)\in
\text{Lie}\left(Sp(2,\R)\ltimes \R^2\right)
\label{mt}
\end{equation} 
of $Sp(2,\R)\ltimes \R^2$. The latter is a group of automorphism of
the Moyal algebra $\A$ and plays for the Moyal plane the same
role as the Euclidean group for $\R^2$ (see e.g.
 \cite{Lizzi:2001fk}). 

In all this section, $S^{\mu\nu}\in\R$ are the components of the matrix $S\in\frak
  sp(2,\R)$, $\kappa^\mu$ of the vector{\footnote{We use the
      same notation for the group $\R^2$ and its Lie algebra.}}
  $\kappa\in\R ^2$, $J^\mu$ of the vector ${\bf J}$ tangent to the curve $M(t)x = e^{tJ}x$ at
$t=0$. More explicitly
\begin{equation}
J^\mu(x) \doteq (Jx)^\mu = S^{\mu\nu}x_\nu + \kappa^\mu {\bf 1}.
\label{eq:244}
\end{equation}
The square norm of ${\bf J}$ and its evaluation on a function $f$ are
respectively denoted by
  \begin{equation}
{\bf J}\cdot {\bf J} = \sum_\mu (J^\mu)^2,\quad    {\bf J}\cdot \nabla f \doteq J^\mu\cdot\partial_\mu f.
  \label{eq:69}
   \end{equation}  
Similarly we write
\begin{equation}
  \label{eq:80}
  {\bf J}\star {\bf J} \doteq \sum_\mu J^\mu\star J^\mu,\quad    {\bf J}\star \nabla f \doteq J^\mu\star\partial_\mu f.
\end{equation}
\subsection{Monge-Kantorovich distance on symplectic orbits}

Let us begin with the commutative case and consider the spectral triple $\left(S(\R^2), \hh, D\right)$ with $\hh$ and
  $D$ as in (\ref{eq:2}) (for $N=1$). The algebra acts by pointwise
  multiplication, so a real function $f$ is in the Lipschitz ball iff
$\norm{\nabla f\cdot\nabla f}_{\infty}\leq
1$. We still call $d_D$ the spectral distance, keeping in mind it coincides with the
Monge-Kantorovich distance.
${\cal S}_0(C_0(\R^2))$ denotes the convex hull of
the Schwartz pure states of $C_0(\R^2)$ (i.e. $f\to \int_{\R^2} \abs{\psi}^2
f dx$ with $\psi\in S(\R^2)$).  For any such state, $\varphi\left({\bf J}\cdot {\bf
    J}\right)$ and $\varphi\left({\bf J}\cdot \nabla f\right)$ are
well defined.
\begin{prop} 
\label{propsymp}
 Let $\varphi$ be a faithful state in ${\cal S}(C_0(\R^2))$ and
 $M(t)=e^{t\tilde{J}}\in Sp(2,\R)\ltimes \R^2$.
Then for any $\tau\in\R^+$
\begin{equation}
\label{Veneziater}
 d_D\left(\varphi, \varphi\circ \alpha_{M(\tau)}\right)\leq \inf_N\,
\int_0^{\tau} dt\; \varphi_{t}\left(\sqrt{{\bf J}\cdot{\bf J}}\right)
\end{equation}
where the infimum is on all $N(t)=e^{tJ}\in Sp(2,\R)\ltimes
\R^2$ such that
$\varphi\circ\alpha_{N(\tau)} = \varphi\circ\alpha_{M(\tau)}$.  \end{prop}
\begin{preuve}
For any real function $f$ in the Lipschitz ball and $N(t)=e^{tJ}$, define $F(t)\doteq
\varphi_t(f)= \varphi\circ\alpha_{N(t)}(f)$. By lemma \ref{lemupper} one has
\begin{equation}
\abs{\frac{dF}{dt}_{\lvert t}}
= \abs{\varphi_t\left(  {\bf J}\cdot \nabla f\right)}.
\label{eq:242}
\end{equation}
The $L^2$-convergence (\ref{eq:127bis})  is guaranteed, for
$J^\mu$ is a linear combination of the coordinates and ${\bf
  1}$. 

Eq.(\ref{Veneziater}) follows if we show that
$\abs{\varphi_t\left( {\bf J} \cdot\nabla f\right)} \leq
\varphi_t\left( \sqrt{{\bf  J}\cdot {\bf J}}\right).$
 Denoting $\phi_t$ the density probability of $\varphi_t$, 
this comes from  Cauchy-Schwarz,
 \begin{align}
   \label{eq:93}
  \abs{\varphi_t({\bf J}\cdot \nabla f)} &=\abs{\int_{\R^2} \!\! dx\; \phi_t\; {\bf J}\cdot
   \nabla f} \leq \int_{\R^2} \!\!dx\; \phi_t\, \abs{{\bf J}\cdot \nabla f}
 \leq \int_{\R^2}\!\! dx\; \phi_t \,\sqrt{{\bf  J}\cdot {\bf J}}\,\sqrt{\nabla
   f\cdot\nabla f},\\
&\nonumber\leq \sup_x \sqrt{\nabla
   f\cdot\nabla f}\int_{\R^2}\!\! dx\; \phi_t\,\ \sqrt{{\bf  J}\cdot {\bf J}}\leq \varphi_t\left( \sqrt{{\bf  J}\cdot {\bf
       J}}\right). \end{align}

\vspace{-.95truecm}\end{preuve}
\newline

In fact, we are dealing with the full fledged time-dependent
Monge-Kantorovich problem and it is known that
to have  equality in (\ref{Veneziater}) in the general case, the infimum should be taken on
\textit{all} continuous and piecewise $C^1$ trajectories $N(t)$ such that $\varphi\circ\alpha_{M(\tau)}
 =\varphi\circ\alpha_{N(\tau)}$  (cf e.g. \cite[\S
 5.1]{Villani:2003fk}).

However, we show in 
proposition \ref{proprot} below that (\ref{Veneziater}) provides the right bound for states
with specific symmetry properties. We stress that in this case the spectral distance does \emph{not}
coincide with the integral of $\varphi_t\left(\sqrt{{\bf J}\cdot{\bf J}}\right)$ along the orbit $M(t)$. 
This already happens with pure states: the orbit $\delta_t \doteq \delta_x\circ\alpha_{M(t)}$ in
$\pa$ identifies with the curve $M(t)x$ in the plane and
$\delta_t\left(\sqrt{{\bf J}\cdot {\bf J}}\right)$ is the norm of the
vector tangent to this curve at $t$ so that,  for instance with $M(t)$ a
rotation, one obtains
\begin{equation}
\int_0^\tau dt \,\delta_t\left(\sqrt{{\bf J}\cdot {\bf
      J}}\right) = \text{length of the arc of angle $\tau$ and
radius $\abs{x}$}\label{eq:175}
\end{equation}
while by (\ref{dgeo}) the spectral distance is the length of the corresponding chord,
\begin{equation}
d_D(\delta_x, \delta_\tau)= d_{\text{geo}}(x, M(\tau)x) =\abs{M(\tau)x
  - x}.
\label{eq:120}
\end{equation}

\subsection{Symplectic orbits on the Moyal plane} 
Let us now consider the orbits of the symplectic group in the Moyal
plane. For any $\varphi\in\s0a$ (defined in remark \ref{remstate}) 
$\varphi\left({\bf J}\star{\bf
    J}\right)$, $\varphi\left(\nabla f \star {\bf J}\right)$ and $\varphi\left({\bf J}\star \nabla f\right)$ are
well defined finite quantities. The same is true for
$\varphi\left({\bf J}\cdot\nabla f\right)$, as shown by the following lemma.

\begin{lem} 
\label{lemmeta}
For any Schwartz function $f$, one has
  \begin{equation} 
{\bf J}\cdot \nabla f =\frac 12\left(\nabla f\star
      {\bf J} +
      {\bf  J}\star \nabla f\right).\label{eq:213}
  \end{equation}
\end{lem}
\begin{preuve}
This follows from $f\cdot x_\mu =\frac
12\left(f\star x_\mu + x_\mu\star f\right)$, see \cite{Figueroa:2001fk}.   
\end{preuve}
\newline

We now give an upper bound to the spectral distance on any symplectic orbit.
\begin{prop} 
\label{propsympmoyal}
 Let $\varphi\in\s0a$ and $\varphi_t=\varphi\circ\alpha_{M(t)}$ its
   orbit under $M(t)=e^{t\tilde J}$. Then $\forall\,\tau\in\R^+$
\begin{equation}
\label{Veneziabis}
 d_D\left(\varphi, \varphi_\tau\right)\leq \inf_N
\int_0^{\tau} dt\; \sqrt{\varphi_{t}\!\left({\bf J}\star {\bf J}\right)}
\end{equation}
 where the infimum is on all $N(t)=e^{tJ}\in Sp(2,\R)\ltimes
\R^2$ such  that $\varphi\circ\alpha_{N(\tau)}=
 \varphi\circ\alpha_{M(\tau)}$.
\end{prop}
\begin{preuve}
As in prop. \ref{propsymp}, the result 
follows if we show that for any $f=f^*$ in the Lipschitz ball,
\begin{equation}
\abs{\varphi_t\left( {\bf J} \cdot\nabla f\right)} \leq
\sqrt{\varphi_t\left({\bf J} \star {\bf J} \right)}.
\label{eq:132}
\end{equation}
The point is that the usual product appears in l.h.s. of (\ref{eq:132}), not the star
one. So a blind application of Cauchy-Schwarz inequality would make the sup norm $\norm{\partial f}_\infty$
arise, which is no longer constrained by the Lipschitz
condition. Fortunately, thanks to 
lemma \ref{lemmeta} we can substitute the argument of $\varphi_t$ with
$\frac 12\left(\nabla f\star {\bf J} +  {\bf J}\star \nabla f\right)$.
We then switch to complex coordinates 
\begin{equation}
\partial_{a=1}\doteq \partial,\;  \partial_{a=2}\doteq \bar\partial,\quad  J^{a=1}\doteq\frac{J^{\mu=1} + i  J^{\mu=2}}{\sqrt 2},\quad
  J^{a=2}\doteq\frac{J^{\mu=1} - i  J^{\mu=2}}{\sqrt 2} = \left(J^{a=1}\right)^*.\label{eq:209}
\end{equation}
One checks that
$\nabla f \star {\bf J}=  \partial_a  f\star J^a$
and  ${\bf J} \star {\bf J} = \sum_a
J^{a*}\star J^a$.
By Cauchy-Schwarz, one thus has
{\footnote{No Einstein summation.}} 
\begin{align}
\nonumber
\abs{\varphi_t\left(\nabla f\star {\bf J} \right)}&\leq \sum_a
\abs{\varphi_t( \partial_a f\star  J^a)}  \leq \sum_a\sqrt{\varphi_t\!\left(\partial_a
    f\! \star\! ( \partial_a f)^*\right)
  \,\varphi_t\!\left(J^{a*}\!\star\! J^a\right)}\,\\
&\leq 
\sqrt{\sum_a\varphi_t\!\left(\partial_a f\!\star(\partial_a f)^* \right)}\sqrt{\sum_a \varphi_t\!\left(J^{a*}\!\star J^a\right)} \leq 
\nonumber
\sqrt{\sum_a\norm{\ll(\partial_a f)}^2}\,\sqrt{\sum_a\varphi_t\!\left(
    J^{a*}\!\star\! J^a\right)}\\
&\label{genova}\leq 
\sqrt{\sum_a\varphi_t\!\left(J^{a*}\!\star\! J^a\right)} =
\sqrt{\varphi_t\left({\bf J}\star {\bf J}\right)}.
\end{align}
A similar equation holds for $\abs{\varphi_t\left( {\bf J}\star \nabla
    f \right)}$. Hence (\ref{eq:132}) and the result.
\end{preuve}

\begin{rem}
1. It is known that the spectral distance on $\sa$ can be infinite
\cite{Cagnache:2009oe}. Proposition \ref{propsymp} guarantees that $d_D(\varphi,
 \varphi\circ\alpha_{M(\tau)})$ is finite for
any $\varphi\in\sa$ and $M(\tau)\in Sp(2,\R)$.

\noindent 2.  The upper bound (\ref{Veneziabis}) is greater than what
could have been expected from
proposition \ref{propsymp}:
\begin{equation}
\sqrt{\varphi_{t}\!\left({\bf J}\star {\bf J}\right)}
~\geq~\varphi_{t}\!\left(\sqrt{{\bf J}\star {\bf J}}\right).
\label{eq:176}
\end{equation}
\end{rem}

When $\varphi$ is invariant under a given symplectic
transformation $S$, the distance on the orbit of its
translated $\varphi_\kappa$ is easily computable.
\begin{prop}
\label{proprot}
Let $\varphi=\varphi\circ\alpha_S$ be a state invariant under the
symplectic transformation $S$. Then 
\begin{equation}
d_D(\varphi_\kappa, \varphi_{\kappa}\circ\alpha_S)=\abs{S\kappa
  -\kappa} \quad\forall\kappa\in\R^2.\label{eq:231}
\end{equation}
  \end{prop}
\begin{preuve}
The result follows from theorem \ref{theomain}, noticing that  the group law of $Sp(2,\R)\ltimes \R^2$ guarantees that the image of
any translated $\varphi_\kappa$ under $\alpha_S$ is still a
translated of $\varphi$. Namely
\begin{equation*}
\varphi_\kappa\circ\alpha_S
=\varphi\circ(\alpha_\kappa\alpha_S)=\varphi\circ\alpha_S\circ
\alpha_{S\kappa} = \varphi\circ\alpha_{S\kappa}
= \varphi_{S\kappa}.\label{eq:210}
 \end{equation*}

\vspace{-.81truecm}
 \end{preuve}
\newline

\noindent This proposition sheds an interesting light on the nature of a quantum point.
In the Euclidean plane $d_D(\delta_x,
\delta_x\circ\alpha_S) = \abs{Sx-x}$. In view of proposition
\ref{proprot} , this is because any $\delta_x$
in $\R^2$ is the translated of the origin $\delta_0$, and the latter is invariant
under any symplectic transformation. This is no longer true on the
Moyal plane: the set
$\ccc(\varphi)\doteq \left\{\varphi_\kappa, \kappa\in\R^2\right\}$ equipped
with the spectral distance is homeomorphic (for the metric topology)
to the Euclidean plane, but unlike $\delta_0$ the ``origin''
$\varphi$ may not be invariant
under $S$. When this happens $\varphi_\kappa\circ\alpha_{S}$ no
longer lies in $\ccc(\varphi)$ and $d_D(\varphi_\kappa,
\varphi_\kappa\circ\alpha_S)$ is no longer the length of a chord.
\newline

Propositions \ref{propsymp} and \ref{propsympmoyal} indicate that in both
the commutative and noncommutative cases the techniques developed in
section \ref{distancetranslationsection} do not straightforwardly
apply to symplectic transformations. Except for
translations, $d_D(\varphi, \varphi\circ\alpha_{M(\tau)})$ is
not necessarily given by the integral of some ``Moyal line-element'' (like $\sqrt{\varphi_t\left({\bf J}\star{\bf
       J}\right)}$ or $\varphi_t\left(\sqrt{{\bf J}\star{\bf
       J}}\right)$) along the orbit $\varphi\circ\alpha_{M(t)}$.
\newpage

\section{Pythagoras relations}
\label{secPyth}
\subsection{Product of spectral triples}
The product of an (even) spectral triple $T_1 = (\A_1, \hh_1, D_1)$
with a spectral triple $T_2 = (\A_2, \hh_2, D_2)$ 
is the spectral triple $T'=(\A', \HH', D')$ with \cite{Connes:1996fu}
\begin{equation}
  \label{eq:104000}
  \A' \doteq \A_1\otimes \A_2,\quad \HH' \doteq \HH_1\otimes \HH_2, \quad D'\doteq
 D_1\ot \I_2 + \Gamma_1\ot D_2,
\end{equation}
where $\Gamma_1$ denotes the chirality of $T_1$, that is a graduation
of $\hh_1$ such that
\begin{equation}
\Gamma_1^2=\ii_1 \text{ is the identity of }\; \hh_1 \; \text{ and }\; [\Gamma_1,
\pi_1(\A_1)]=0.
\label{eq:181}
\end{equation}

\begin{figure}[h*]
\begin{center}
\vspace{-8truecm}
\mbox{\rotatebox{0}{\scalebox{.65}{\hspace{1truecm}\includegraphics{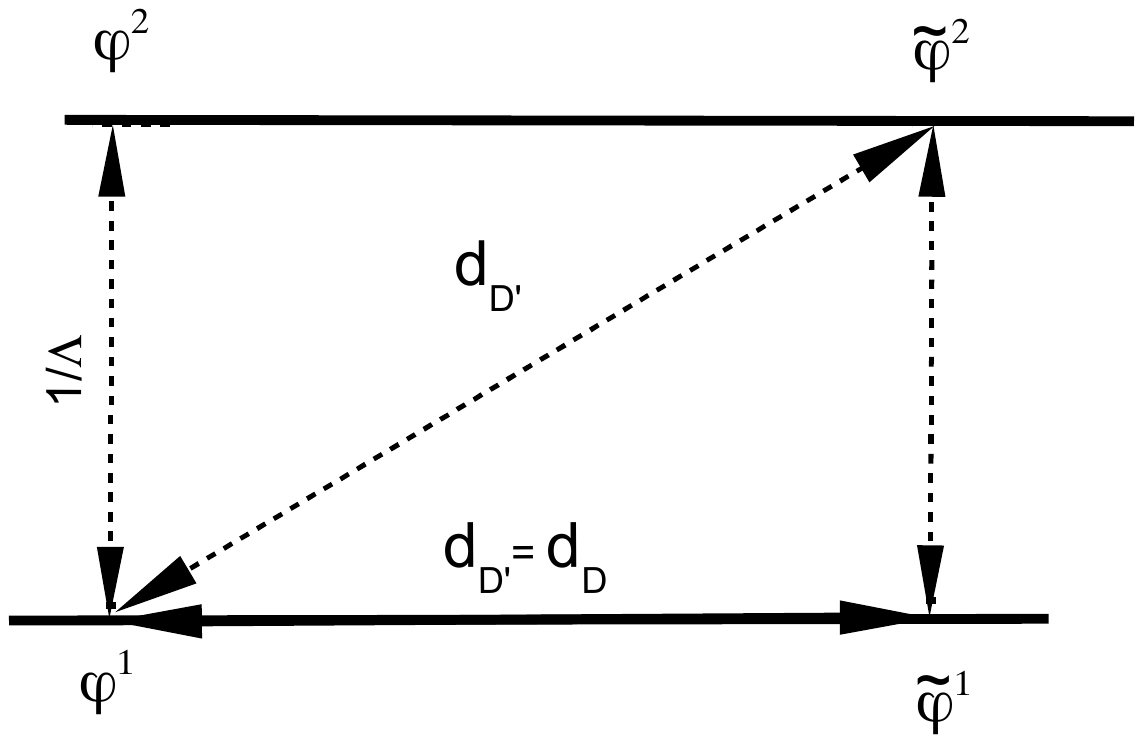}}}}
\end{center}
\vspace{-6.5truecm}
\caption{\small The spectral distance on $\qq(\varphi)$: $\varphi$ is any
  state of the Moyal algebra and $\tilde\varphi\in\ccc(\varphi)$ any of its
  translated. The superscript $1,2$ corresponds to the pure state of
  $\C^2$.  On each copy of $\ccc(\varphi)$ the spectral distance
  $d_{D'}$ coincides with the distance $d_D$ on the Moyal plane, and
  the distance between two copies $\varphi^1, \varphi^2$ of the same
  state $\varphi$ coincides with
  the distance $d_{D_2}$ on $\C^2$. The
  distance between the two sheets is given by Pythagoras theorem.}
\label{coherentfig}
\end{figure}

Any pair of states $\left(\varphi_1\in\ss(\A_1),
  \varphi_2\in\ss(\A_2)\right)$ defines state of
$\A'$ by $a_1\otimes a_2 \to \varphi_1(a_1) \varphi_2(a_2)$.
Any state of $\A'$ that can be written in this way is said
\emph{separable}.
Consider two separable states either of the form 
$\varphi_1\otimes \varphi_2, \,\varphi_1\otimes \tilde\varphi_2$
or
$\varphi_1\otimes \varphi_2, \,\tilde \varphi_1\otimes
\varphi_2$ and  make the extra-assumption that both spectral triples $T_i$, $i=1,2$,
are unital and
non-degenerate (that is $\pi_i({\bf 1})$ is the identity of $\hh_i$). Then
 the
spectral distance $d_{D'}$ of $T'$
reduces either to the distance $d_{D_1}$ of $T_1$ or $d_{D_2}$ of
$T_2$ \cite{Martinetti:2002ij,Martinetti:2001fk}:
 \begin{align}
\label{distext}
&d_{D'}( \varphi_1\otimes \varphi_2, \;\tilde \varphi_1\otimes \varphi_2) = d_{D_1}(\varphi_1, \tilde \varphi_1),\\
\label{distint}&d_{D'}(\varphi_1\otimes \varphi_2, \;\varphi_1\otimes \tilde\varphi_2 ) =
d_{D_2}(\varphi_2, \;\tilde\varphi_2 ).
\end{align} 
Furthermore, in case $T_1$ is the canonical (commutative) spectral
triple of a compact spin manifold $\mm$ and $T_2$ is the canonical spectral triple on $\C^2$ given by
\begin{equation}
\label{triplinterne}
\A_2=\C^2,\quad \hh_2=\C^2,\quad D_2 \doteq \left( 
\begin{array}{cc} 0& \Lambda \\ \overline{\Lambda}&0 \end{array}
  \right)\quad 
\end{equation}
where $\Lambda$ is a constant complex parameter and $\C^2$ acts on itself as
 \begin{equation}
   \label{eq:110}
   \pi_2(z^1, z^2)\doteq \left(\begin{array}{cc} z^1 & 0 \\ 0&
       z^2\end{array}\right) \quad\quad z^1, z^2\in\C,
 \end{equation}
then Pythagoras equality
holds true between pure states of $\A'$ (which turn out to be all
separable),
that is \cite{Martinetti:2002ij,Martinetti:2001fk}
\begin{equation}
  \label{eq:161}
 d_{D'}(\delta_x^1, \delta_y^2) = \sqrt{d_{D'}^2(\delta_x^1,
   \delta_y^1) + d_{D'}^2(\delta_y^1, \delta_y^2)}
\end{equation}
where $\delta_x^i=(\omega_x, \delta^i)$ with $\delta_x\in\pp(C_0^\infty(\M)$ and $\delta^i\in\pp(\aa_1)$ is one of the
two pure states of $\C^2$,
\begin{equation}
\delta^i(z^1, z^2) = z^i\quad\quad i=1,2.
\label{eq:173}
\end{equation}
\begin{rem}A similar result holds for $\A_2=\C \oplus \hhh \oplus M_3(\C)$ 
the finite dimensional algebra of
the standard model. In fact one shows \cite{Martinetti:2001fk,
  Martinetti:2002ij} that with this choice of $\A_2$, the computation
of $d_{D'}$ between separable states is equivalent to the $\A_2= \C^2$ case.
Indeed, the requirements imposed by the experimental data on the Dirac
operator make all the distance between states of $M_3(\C)$ infinite,
while the algebra of quaternion $\hhh$  has only one state and thus,
from our perspective, behaves like the algebra of complex numbers.
\end{rem}
We prove below that the same Pythagoras equality holds in the
Moyal plane - modulo minimal unitalization - on any set of separable
states of $\A' =\A\otimes \C^2$ of the form (see figure \ref{coherentfig})
\begin{equation}
\qq(\varphi) \doteq\ccc(\varphi)\times \ccc(\varphi) =\left\{\left(\ccc(\varphi), \delta^i\right), i=1,2\right\}.
\label{eq:164}
\end{equation}
To do so, we first establish some Pythagoras inequality
(\ref{eq:108q}),
which holds true for the product of any arbitrary unital, non-degenerate spectral triple
$T_1$ by $\C^2$. 

\subsection{Pythagoras inequalities}
\label{Pythineq}
In this section $\A' = \A_1 \otimes \C^2\simeq \A_1\oplus \A_1$ has
generic element $a'= (f,g)$ with
 $f,g\in\A_1$. We consider the subspace $\qq(\aa')$ of $\ss(\A')$ consisting in pairs
\begin{equation}
\varphi^i \doteq
(\varphi, \delta^i)
\label{eq:95xx}
\end{equation}
where
$\varphi\in{\mathcal S}(\A_1)$.
$\qq(\A') = \underset{\varphi\in\ss(\A_1)}\cup \qq(\varphi)$
 is the disjoint union of two copies of
$\mathcal{S}(\A_1)$. The evaluation on an element of $\A'$ reads
\begin{equation}
  \label{eq:141}
  \varphi^1(a') = \varphi(f), \quad\varphi^2(a') = \varphi(g).
\end{equation}
Eqs. (\ref{distext}) and (\ref{distint}) become
\begin{align}
\label{distextbis}
&d_{D'}(\varphi^i, \,\tilde \varphi^i) = d_{D_1}(\varphi, \tilde \varphi),\\
\label{distintbis}&d_{D'}(\varphi^1, \,\varphi^2) =
d_{D_2}(\delta^1,\delta^2)= \abs{\Lambda}^{-1}.
\end{align}
Note that $\A_2$ being commutative,  pure states of $\A'$
are pairs \cite{Sakai1971}
$\omega^i \doteq \left(\omega\in\pp\left(\A_1\right), \delta^i\right)$,
so that 
\begin{equation}
  \label{eq:153}
  \pp(\A') \subset \qq(\A')\subset \ss(\A').
\end{equation}
\begin{lem}
\label{normnormpyth}
 Let $(\A_1, \HH_1, D_1)$ be any spectral triple. For any states $\varphi,
 \tilde\varphi\in\ss(\A_1)$ and any $a$ in $\bb_{\text{Lip}}(T_1)$ that does not commute
 with $D_1$, one has
  $\abs{\varphi(a) - \tilde\varphi(a)}\leq \norm{[D_1,\pi_1(a)]} d_{D_1}(\varphi, \tilde\varphi).$
\end{lem}
\noindent\begin{preuve}
Let $\tilde a \doteq \frac a{\norm{[D_1,\pi(a)]}}$. Then
 $\norm{[D_1,\pi_1(\tilde a)]}=1$. Hence the result by definition of $d_{D_1}$.
 \end{preuve}

\begin{prop}\label{pyth}
 Let $(\A', \HH', D')$ be the product (\ref{eq:104000}) of an arbitrary
 even, unital and non-degenerate spectral triple $T_1$ with the spectral triple $T_2$ (\ref{triplinterne}) of $\C^2$. For any $\varphi^1, \tilde \varphi^2
 \in\qq(\A')$,
  \begin{equation}
    \label{eq:108q}
   \sqrt{d_{D'}^2(\varphi^1,\tilde\varphi^1) + d^2_{D'}(\tilde\varphi^1, \tilde \varphi^2)} \leq
    d_{D'}(\varphi^1, \tilde \varphi^2) \leq 
\sqrt 2  \sqrt{d_{D'}^2(\varphi^1,\tilde\varphi^1) + d^2_{D'}(\tilde\varphi^1, \tilde \varphi^2)} .
  \end{equation}
\end{prop}
\begin{preuve} 
$a'= (f,g)$ is represented as
  \begin{equation}
    \label{eq:111q}
    \pi'(a') = \left(\begin{array}{cc} \pi_1(f)& 0\\ 0& \pi_1(g)\end{array}\right).
  \end{equation}
The Dirac operator $D'$ acts as 
\begin{equation}
  \label{eq:100c}
  D' = \left( \begin{array}{cc}
D_1 &\Lambda\Gamma_1 \\ \bar \Lambda\Gamma_1 & D_1
\end{array}
\right)
\end{equation}
so that, by (\ref{eq:181}),
\begin{equation}
  \label{eq:100d}
 [D', \pi'(a)] = \left( \begin{array}{cc}
[D_1, \pi_1(f)] & \Lambda\Gamma_1 \,\pi_1(f- g) \\ \bar \Lambda \Gamma_1\,
\pi_1(g - f)& [D_1, \pi_1(g)]
\end{array}
\right).
\end{equation}

 Define the subset
 \begin{equation}
\A'\supset \bb \doteq \left\{ (f,g)=(f^*, g^*) \in
   \A', \, f-g = \lambda {\bf 1} \text{ for some }
   \lambda\in\R^+\right\}
\label{eq:165}
 \end{equation}
 with {\bf 1} the unit of $\A_1$, and let
\begin{equation}
  \label{eq:98q}
  d_\bb(\varphi^1, \tilde\varphi^2) \doteq \suup{b\in\bb}\left\{ \abs{\varphi^1(b)
    -\tilde\varphi^2(b)}, \, \norm{[D',\pi'(b)]}\leq 1\right\}.
\end{equation}
Obviously $d_{D'}(\varphi^1, \tilde\varphi^2) \geq
d_\bb(\varphi^1, \tilde\varphi^2)$, so the l.h.s. of
(\ref{eq:108q}) follows if we show that 
  \begin{equation}
\label{eqtest}
  d_\bb(\varphi^1, \tilde\varphi^2) = \sqrt{ d_{D'}^2(\varphi^1,\tilde\varphi^1) + d_{D'}^2(\tilde\varphi^1, \tilde\varphi^2)}. 
\end{equation}

To this aim, let us fix $\varphi, \tilde\varphi$ in $\ss(\A_1)$, and consider
\begin{equation}
b=(f, f +
\lambda {\bf 1})\in\bb\cap\bb_{\text{Lip}}(T').
\label{eq:166q}
\end{equation}
Noticing that $[D_1, \pi_1(f)]^* =
 -[D_1, \pi_1(f)]$, (\ref{eq:100d}) yields
\begin{equation}
  \label{eq:100e}
 [D', \pi'(b)][D', \pi'(b)]^* = \left([D_1, \pi_1(f)][D_1, \pi_1(f)]^*
   + \lambda^2\abs{\Lambda}^2 \I_1\right) \otimes \I_2.
\end{equation}
For any positive element in a unital
$C^*$-algebra and $\lambda\in\R^+$, $\norm{a+ \lambda {\bf 1}} = \norm{a} +
\lambda$, hence
\setcounter{footnote}{3}
\begin{equation}
  \label{eq:106q}
  1\geq \norm{[D', \pi'(b)]} =  \sqrt{\norm{[D_1, \pi_1(f)]}^2
    + \lambda^2\abs{\Lambda}^2 }.
\end{equation}
Furthermore, for any $b\in\bb$  lemma \ref{normnormpyth} yields
\begin{equation}
  \label{eq:99q}
\abs{\varphi^1(b) -\tilde\varphi^2(b)} = 
\abs{\varphi(f) -\tilde\varphi(f) + \lambda} \leq
d_{D_1}(\varphi, \tilde\varphi) 
\norm{[D_1,
    \pi_1(f)]}+ \lambda.
 \end{equation}
Therefore, using  (\ref{eq:106q}) as $\norm{[D_1,\pi_1(f)]}\leq \sqrt{1-
  \lambda^2 \abs{\Lambda}^2}$, one obtains 
\begin{equation}
  \label{eq:113q}
  d_\bb(\varphi^1, \tilde\varphi^2)  \leq
  \sup_{\lambda\in\R^+} F(\lambda) \quad \text{ with }\quad F(\lambda)\doteq \sqrt{1- \lambda^2 \abs{\Lambda}^2}d_{1} + \lambda
\end{equation}
where we write $d_1$ instead of
$d_{D_1}(\varphi, \varphi')$. The function $ F$ reaches its maximum on $\R^+$, 
\begin{equation}
  \label{eq:115q}
  F(\lambda_{\text{max}}) = \sqrt{\frac 1{\abs{\Lambda}^2} + d_1^2},
\end{equation}
for
\begin{equation}
\lambda_{\text{max}} = \frac{1}{\abs{\Lambda}\sqrt{d_1^2\abs{\Lambda}^2 + 1}}.\label{eq:1170}
\end{equation}
This upper bound in the spectral distance formula is attained by $
b=(f, f+\lambda_\text{max}{\bf 1})$ with
\begin{equation}
  \label{eq:118q}
  f=\frac{\abs{\Lambda}d_1}{\sqrt{\abs{\Lambda}^2d_1^2 +1}} f_1, 
\end{equation}
where $f_1$ is the element that attains the supremum in the
computation of $d_1$. Therefore
\begin{equation}
d_\bb(\varphi^1, \tilde\varphi^2)
= \sqrt{\frac 1{\abs{\Lambda}^2} + d_1^2}
\label{eq:167q}
\end{equation}
which, by (\ref{distextbis}),~(\ref{distintbis}),
is nothing but
(\ref{eqtest}).
Hence the l.h.s. of (\ref{eq:108q}). 

The r.h.s. follows from (\ref{distextbis}) and (\ref{distintbis}) by the triangle inequality,
\begin{equation}
  \label{eq:119q}
  d_{D'}(\varphi^1, \tilde\varphi^2) \leq d_{D'}(\varphi^1, \tilde\varphi^1) + d_{D'}(\tilde\varphi^1,\tilde\varphi^2),
\end{equation}
together with $(a+b)^2 \leq 2a^2 + 2b^2$.
\end{preuve}
\newline

\noindent  This result is extended in \cite{DAndrea:2012fk} to the distance between separable
  states of $\A_1 \otimes \A_2$ in the product of any two unital
  non-degenerate spectral triples (that is for $\A_2$ non-necessarily
  equal to $\C^2$). Below, in
  the particular case of the product of the Moyal plane by $\C^2$, we
  obtain a stronger result consisting in a Pythagoras equality for
  states $\varphi^1, \tilde\varphi^2$ with
  $\tilde\varphi\in\ccc(\varphi)$ a translated of $\varphi$.

\subsection{Pythagoras equality}
\label{pythagore}

Since the Moyal algebra $\A$ has no unit, in order to apply proposition \ref{pyth} above
we shall work with the (minimal) unitization $\A^+$ of $\A$, that is
\begin{equation}
\A^+ =  S(\R^2, \star)
\oplus \C
\label{eq:179}
\end{equation}
as a vector space, with product $(f,\lambda).(g,\tilde\lambda)
= (f g + \lambda g+\tilde\lambda f, \lambda \tilde\lambda)$ and unit
${\bf 1}= (0,1)$.  The left-regular representation and the
representation $\pi$ of the spectral triple (\ref{eq:184}) extend to
$\A^+$ as
\begin{align}
  \label{eq:180}
  &\ll(f_\lambda)\doteq \ll(f) + \lambda\I,\\ 
   \label{eq:183}
  &\pi (f_\lambda) = \ll(f_\lambda)\otimes \I_2 = \pi(f) +
  \lambda\ii\otimes\ii_2 \quad \text{ where } f_\lambda \doteq (f,\lambda).
\end{align}
Any state $\varphi\in\sa$ linearly extends to a state in $\ss(\A^+)$ by 
setting $\varphi({\bf 1}) = 1$. 

One should be careful that $T^+ = (\A^+, \hh, D)$ with $\hh$ and $D$ given by (\ref{eq:41}) and (\ref{eq:33})  is \emph{not} a spectral
triple, for $\pi(a)(D-\lambda\I)^{-1}$ is not compact for $a= \bf
1$. However this does not prevent us to consider the associated
spectral distance, since the compact resolvent condition does not
enter formula (\ref{eq:22}). Actually Rieffel considers such
generalizations of Connes formula in \cite{Rieffel:1999ec}, using instead of (\ref{eq:53})  any
seminorm $L$ on $\A$. 
Within this framework,  we call $T^+$ the \emph{unital triple} of
the Moyal plane and still call $d_D$ the spectral distance.

Switching from $T$ to $T^+$ has no incidence on the spectral distance
on $\sa$.
\begin{lem}
\label{nonunit}
Let $(\A_1, \HH_1, D_1)$ be a non-unital, non-degenerate spectral triple. For any $\varphi, \tilde\varphi\in \ss(\A_1)$, 
\begin{equation}
d_{D_1}(\varphi, \tilde\varphi) =\suup{a\in \A_1^+}\left\{ \abs{\varphi(a)
  -\tilde\varphi (a)},\; \norm{[D_1,\pi_1(a)]} \leq 1\right\}
\end{equation}
where $\pi_1$ is extended to $\A_1^+$ by setting $\pi_1({\bf 1}) = \I_1$.
\end{lem}
\begin{preuve} One has
$\varphi((a,\lambda)) = \varphi(a) + \lambda$, so that
$\varphi((a,\lambda))-\tilde\varphi((a,\lambda)) =
\varphi(a)-\tilde\varphi(a)$ does not depend on
$\lambda$. Similarly
$[D_1, \pi_1(a,\lambda)] = [D_1, \pi_1(a) + \lambda \I_1]
= [D_1, \pi_1(a)]\label{eq:182}$
does not depend on $\lambda$. Hence it is equivalent to
look for the supremum on $\A_1$ or on $\A_1^+$.
\end{preuve}
\newline

We now consider the product $T'$ of  the unital
triple $T^+$ of the Moyal plane with the spectral triple $T_2$ on
$\C^2$ given in  (\ref{triplinterne}), namely
\begin{equation}
\A' = \A^+\ot \C^2,\quad  \hh' = \hh\ot \C^2,\quad  D' = D\ot \ii_2 +
\Gamma \ot D_2.
\label{eq:185}
\end{equation}
The graduation $\Gamma$ is the third Pauli
matrix $\sigma_3$. 
We begin with a technical lemma regarding
the Lipschitz ball.

\begin{lem}
  \label{lipd'}
For any $a' = \left(f_{\lambda}, g_{\tilde{\lambda}}\right)$ in $\bb_{\text{Lip}}(T')$, with
$f_{\lambda}, g_{\tilde{\lambda}}\in\A^+$, one has
\begin{align}
\label{normf}
  \norm{\ll \left(\partial f \right)^*\ll \left(\partial f \right) + \frac{\abs{\Lambda}^2}2
\ll\left(g_{\tilde{\lambda}}-f_{\lambda} \right)^*\ll \left(g_{\tilde{\lambda}}-f_{\lambda} \right)}\leq &\frac 1{2},\\
\label{normg}
   \norm{\ll\left(\partial g\right)^*\ll\left(\partial g\right) + \frac{\abs{\Lambda}^2}2
\ll\left(g_{\tilde{\lambda}}-f_{\lambda}\right)^*\ll\left(g_{\tilde{\lambda}}-f_{\lambda}\right)}\leq &\frac 1{2},
\end{align}
and similarly for $\bar\partial$.
\end{lem}
\begin{preuve}
From (\ref{eq:100d}), one gets
\begin{equation}
  \label{eq:145}
  [D',\pi'(a')] =\left(\begin{array}{cc}
[D,\pi( f_\lambda)] & \Gamma\Lambda\pi(g_{\tilde\lambda} - f_\lambda)\\
 \Gamma\bar\Lambda\pi(f_\lambda-g_{\tilde\lambda} )&[D,\pi(g_{\tilde\lambda})]
\end{array}\right).
\end{equation}
Multiplying on the left and right by the diagonal matrices of norm $1$
\begin{equation}
\Delta_L\doteq \text{diag}(\ii, 0),\quad\Delta_R\doteq \text{diag}(\ii,
\Gamma),
\label{eq:159}
\end{equation}
one obtains
\begin{equation}
  \label{eq:160}
 \norm{\Delta_L[D',\pi'(a')]\Delta_R} \leq \norm{[D',\pi'(a')]}\leq 1
\end{equation}
where, using the properties (\ref{eq:181}) of $\Gamma$,
\begin{equation}
  \label{eq:157}
  \Delta_L[D',\pi'(a')]\Delta_R =\left(\begin{array}{cc}
[D,\pi(f_\lambda)] & \Lambda\pi(g_{\tilde\lambda}-f_\lambda)\\
0&0
\end{array}\right)
\end{equation}
has norm
\begin{equation}
  \label{eq:158}
  \norm{\Delta_L[D',\pi'(a')]\Delta_R}^2 =
  \norm{[D,\pi(f_\lambda)]^*[D,\pi(f_\lambda)] +
    \abs{\Lambda}^2\pi(g_{\tilde\lambda}-f_\lambda)^*\pi(g_{\tilde\lambda}-f_\lambda)}.
\end{equation}

Let us evaluate the various terms of the equation above.  On the one hand, 
\begin{equation}
  [D,\pi(f_\lambda)] = [D, \pi(f) + \lambda \ii\ot\ii_2]] = [D,\pi(f)].
\end{equation}
so that  by (\ref{eq:34}l
\begin{equation}
[D,\pi(f_\lambda)]^*[D,\pi(f_\lambda)]  =
2\left(
\begin{array}{cc}
\ll(\partial  f)^*\ll(\partial f) & 0 \\
0&\ll(\bar\partial  f)^*\ll(\bar\partial f)
\end{array} \right).
\end{equation}
On the other hand,
\begin{equation}
\label{eq:156} 
 \pi(g_{\tilde\lambda}-f_\lambda) = \ll(g_{\tilde\lambda}-f_\lambda) \otimes \I_2.
\end{equation}
Consequently  $\norm{\Delta_L[D',\pi'(a')]\Delta_R}^2 = \text{max}\left\{
 \norm{A_\partial}, \norm{A_{\bar\partial}}\right\}$ where
 \begin{equation}
  \label{eq:167}
A_\partial \doteq 2\ll\left(\partial f\right)^*\ll\left(\partial f\right)
    +\abs{\Lambda}^2\ll\left(g_{\tilde{\lambda}}- f_{\lambda}\right)^*\ll\left(g_{\tilde{\lambda}}-f_{\lambda}\right),
\end{equation}
and a similar definition for $A_{\bar\partial}$.
Hence~(\ref{normf}). Eq. (\ref{normg}) is obtained in a similar
manner, using $\Delta_L \!\!=\!\! \text{diag}(0,\ii)$.
\end{preuve}
\newline

We now show the second main result of this paper, namely Pythagoras
theorem on the set of states $\qq(\varphi)$ defined in (\ref{eq:164}).
\begin{thm}
\label{theopyth}
For any $\varphi\in\sa$ and $\kappa\in\R^2$, one has
\begin{equation}
  \label{eq:155}
  d_{D'}(\varphi^1, \varphi_\kappa^2) = \sqrt{d_{D'}^2(\varphi^1, \varphi^1_\kappa) + d^2_{D'}(\varphi^1_\kappa,
      \varphi^2_\kappa)} = 
\sqrt{\abs{\kappa}^2 + \abs{\Lambda}^{-2}}.
\end{equation}
\end{thm}
\begin{preuve} We first show that 
\begin{equation}
d_{D'}(\varphi^1, \varphi_\kappa^2)\leq
\sqrt{\abs{\Lambda}^{-2} + \abs{\kappa}^2},
\label{eq:154}
  \end{equation}
using a similar procedure as in proposition
\ref{borne}.  Let
us fix $\varphi\in\sa$, $\kappa\in\C\simeq\R^2$. For any $\A'\ni
a'=(f_{\lambda},g_{\tilde{\lambda}})$
define
\begin{equation}
  \label{eq:136}
 F(u,v) \doteq
 \varphi_{u\kappa}\left(\left(1-v\abs{\Lambda}\right)f_{\lambda} + v\abs{\Lambda}g_{\tilde{\lambda}}\right)
\end{equation}
with $u\in [0,1], v\in[0,\abs{\Lambda}^{-1}]$ and $\varphi_{u\kappa}= \varphi\circ\alpha_{u\kappa}$
the $u\kappa$-translated of $\varphi$ defined in (\ref{eq:104}). One has
\begin{align}
  \label{eq:140}
  F(0,0) =  \varphi(f_{\lambda}) = \varphi^1(a'),\\
 F(u,\abs{\Lambda}^{-1}) =  \varphi_{u\kappa}(g_{\tilde{\lambda}}) = \varphi_{u\kappa}^2(a').
\end{align}
Viewing $F$ as a real function $F(c(t))$ on $\R^2$, where $c(t)= (u(t),
v(t))$ denote a curve in $\R^2$ such that $u(0)= v(0) = 0$, $u(1)=1, 
v(1)=\abs{\Lambda}^{-1}$, one obtains
\begin{equation}
  \label{eq:143}
  \abs{\varphi_\kappa^2(a') - \varphi^1(a')} = \abs{F(c(1)) - F(c(0))}\leq
  \int_0^1 \abs{\frac{dF}{dt}_{\lvert t}} dt.    
\end{equation}
Now fix $c(t) = \left(u \left(t\right) =t, v
  \left(t\right)=\abs{\Lambda}^{-1}t\right)$. The derivative of $F$
along it is 
\begin{equation}
  \label{eq:1500}
\frac{dF}{dt}_{\lvert t} =  \frac{\partial F}{\partial u}_{\lvert t} + 
\abs{\Lambda}^{-1}\frac{\partial F}{\partial v}_{\lvert t}.
\end{equation}
Let us define $K^0 \doteq
\abs{\Lambda}^{-1}$ and
\begin{equation}
\label{derzero}
  \partial_0 F \doteq \frac{\partial F}{\partial v} = \abs{\Lambda}\varphi_{u\kappa}\left(g_{\tilde{\lambda}} -f_{\lambda}\right).
\end{equation}
For $a=\mu=1,2,$ we define as well $K^a = \tilde\kappa^\mu$ and
$\partial_a  = \tilde\partial_\mu$ as
in proposition \ref{borne}, so that by lemma \ref{secup} and
eq.~(\ref{eq:67}) one has $\frac{\partial
  F}{\partial u} = \tilde\kappa^\mu\tilde\partial_\mu F = \sum_{K=1}^2
K^a \partial_a F$. Hence
\begin{equation}
  \label{eq:150}
\frac{dF}{dt}_{\lvert t} = K^a \partial_a F _{\lvert t}
\end{equation}
and by Cauchy-Schwarz one obtains
\begin{equation}
  \label{eq:146}
  \abs{  \frac{dF}{dt}_{\lvert t} }\leq \sqrt{ \abs{\Lambda}^{-2}+
    \abs{\kappa}^2}\;\sqrt{\sum_{a=0}^2  \abs{\partial_a F _{\lvert t}}^2}. 
\end{equation}

Using $\tilde \partial_\mu f_\lambda =
\tilde\partial_\mu f$, $\tilde\partial_\mu g_{\tilde{\lambda}} = \tilde\partial_\mu
g$, we have for $a=1,2$
\begin{align}
\partial_a F = \varphi_{u\kappa}\left(
  \left(1-v  \abs{\Lambda}\right)\tilde\partial_\mu f + v  \abs{\Lambda}\tilde\partial_\mu g\right).
\label{eq:dermu}
\end{align} To lighten notations, let us write
$\tilde f_\mu(t)\doteq \varphi_{t\kappa}(\tilde\partial_\mu f)$
and
 similarly for $g$. Eq. (\ref{eq:dermu}) yields
 \begin{equation}
    \label{eq:148}
\abs{\partial_a F _{\lvert t}}^2 =
 v^2  \abs{\Lambda}^{2}  \left( \tilde g_\mu(t) - \tilde f_\mu(t)\right)^2 +
2v  \abs{\Lambda}\tilde f_\mu(t)\left(\tilde g_\mu(t) - \tilde f_\mu(t)\right) +
\tilde f_\mu^2(t).
  \end{equation}
As a function of $v\in[0, \abs{\Lambda}^{-1}]$, this is a parabola with positive leading
coefficient, hence it is maximum either at $v=0$ or $v=
\abs{\Lambda}^{-1}$. Therefore for $a=1,2$, one has
\begin{equation}
  \label{eq:149}
\abs{\partial_a F _{\lvert t}}\leq
\text{max}\left\{\abs{\tilde {f_\mu}(t)}, \,
  \abs{\tilde g_\mu(t)} \right\} \doteq  \abs{\varphi_{t\kappa}(h_{t\mu})},
\end{equation}
where $h_{t\mu} = \tilde\partial_\mu f $ or $\tilde\partial_\mu g$ is a blind notation to denote the
maximum. It is important to stress that,  at fixed $t$, nothing guarantees that $h_{t1}$ and
 $h_{t2}$ should be given by the same function. One may have
 $h_{t1}=\tilde\partial_1 f$ while $h_{t2}=\tilde\partial_2 g$. As well,
for the same index $\mu$, nothing forbids
 the maximums  at $t_1$ and $t_2\neq t_1$ to  be given by
distinct functions: $\tilde
h_{t_1\mu} = \tilde\partial_\mu f$ and $ h_{t_2\mu}=\tilde\partial_\mu
g$. In any case, since for any state $\varphi$ of a unital $C^*$-algebra
$\abs{\varphi(a)}^2 \leq \varphi(a^*a)$
\cite[Prop. 4.3.1]{Kadison19861}, one gets
\begin{align}
  \label{eq:151}
   \abs{\partial_a F _{\lvert t}}^2 &\leq
   \varphi_{t\kappa}(h_{t\mu}^*\star h_{t\mu}), \text{ for } a=1,2;\\
 \abs{\partial_0 F _{\lvert t}}^2&\leq
 \abs{\Lambda}^{2}\varphi_{t\kappa}\left(  \left(g_{\tilde{\lambda}}-f_{\lambda}\right)^*\star\left(g_{\tilde{\lambda}}-f_{\lambda}\right)\right).
\end{align}
Therefore
\begin{align}
  \label{eq:152}
  \sum_{a=0}^2  \abs{\partial_a F _{\lvert t}}^2 & \leq  \sum_{\mu=1}^2
    \varphi_{t\kappa}\left( h_{t\mu}^*\star
        h_{t\mu} + \frac{\abs{\Lambda}^{2}}2 (g_{\tilde{\lambda}}-f_{\lambda})^*\star(g_{\tilde{\lambda}}-f_{\lambda})\right),\\
& \leq  \sum_{\mu=1}^2
   \norm{\ll\left(h_{t\mu}\right)^*\ll\left(h_{t\mu}\right)+ \frac{
       \abs{\Lambda}^{2}}2 \ll(g_{\tilde{\lambda}}-f_{\lambda})^*\ll(g_{\tilde{\lambda}}-f_{\lambda})} \leq 1
\end{align}
by lemma \ref{lipd'}. Hence (\ref{eq:154}).

The theorem then follows from proposition \ref{pyth}, together with
(\ref{distextbis})  and (\ref{distintbis}).
\end{preuve}
\newline

\begin{rem} The proof of
 theorem \ref{theopyth} relies on the observation that the orbit $\varphi_{t\kappa}\in\ss(\A)$  of $\varphi$ under the action of
the translation group - which allows to define the function $F$ in
(\ref{eq:136}) - is a
 geodesic for the spectral distance, namely 
  \begin{equation}
    \label{eq:128}
    d_D\left(\varphi_{t_0\kappa}, \varphi_{t_1\kappa}\right) = \abs{t_0 - t_1}\,
    d_D(\varphi, \varphi_\kappa) \qquad \forall\; 0\leq t_0 \leq t_1\leq 1.
  \end{equation}
Said differently, the spectral
 distance is intrinsic on $\ccc(\varphi)$, which makes $\left(\ccc(\varphi), d_D\right)$ a  path metric
space \cite{Gromov:1999fk}. This is what makes the improvement from Pythagoras inequalities to
 Pythagoras theorem possible: on a subspace of $\ss(\A)$ which is not
path metric, one should not expect Pythagoras theorem to hold
true.
\end{rem}

\section{Applications}
\label{secdfr}

\subsection{Coherent states and classical limit}
\label{coherentsection}

A coherent state of the Moyal algebra $\A$ is a pure state 
\begin{equation}
  \label{eq:3}
 \omega_\kappa( f) \doteq  \scl{\kappa}{\pi_S(f)\kappa} \quad
  \forall f\in\A
\end{equation}
where $\ket{\kappa}\in L^2(\R)$, $\norm{\kappa}_{L^2(\R)} =1$, is a solution of 
\begin{equation}
  \label{eq:4}
\frak a \ket{\kappa} = \kappa\ket{\kappa} \quad \kappa\in\C.
\end{equation}
Equivalently, $\omega_\kappa$ is the translated of the ground
state
\begin{equation}
\omega_0(\cdot) = \scl{h_0}{\pi_S(\cdot) h_0}
\label{eq:211}
\end{equation}
of the
quantum harmonic oscillator, with translation
$\sqrt{2}\kappa$:
\begin{equation}
  \omega_{\kappa}(f) = \omega_0\circ \alpha_{\sqrt{2}\kappa}(f). 
\label{eq:74}
\end{equation}
The equivalence between (\ref{eq:4}) and (\ref{eq:74}) follows from the unitary implementation (\ref{eq:82}) of translations, 
noticing that 
$\frak u_{\sqrt{2}\kappa}\, h_0 = \ket{\kappa}$. To prove this last
equation, it is convenient to use the development of $\ket{\kappa}$ in the
harmonic oscillator eigenstates basis
  \begin{equation}
    \label{eq:7}
 \ket{\kappa} = \sum_{m\in\N} c^\kappa_m h_m,\quad c^\kappa_m=
e^{-\frac{\abs{\kappa}^2}{2\theta}}\frac{\kappa^m}{\sqrt{m!\theta^m}}
  \end{equation}

\begin{prop}
\label{propcoherent0}
  Let $\omega_\kappa, \omega_{\tilde\kappa}$ be any two coherent
  states of the Moyal algebra, then
  \begin{equation}
    \label{eq:114}
    d_D(\omega_\kappa, \omega_{\tilde\kappa}) =
    \sqrt{2}\abs{\tilde\kappa - \kappa}.
  \end{equation}
On the orbit of a coherent state under rotation, $R\in SO(2)$, the distance is
the length of the chord
\begin{equation}
d_D(\omega_\kappa, \omega_{\kappa}\circ\alpha_R)=\sqrt 2\abs{R\kappa - \kappa}.\label{eq:229}
\end{equation}
  \end{prop}
\begin{preuve}
Eq. (\ref{eq:114}) follows from theorem \ref{theomain} and the
definition (\ref{eq:74}) of a coherent state.
 Eq. (\ref{eq:229}) is an application of proposition \ref{proprot},
 noticing  that the ground state $\omega_0$ is
rotation-invariant because $h_0$ in (\ref{eq:211}) is a
Gaussian centered at the origin.
\end{preuve}
\newline

The coherent state $\omega_\kappa$ reproduces the
behaviour of a classical  harmonic oscillator
with amplitude of oscillation
$\abs{\kappa}$ and phase $\text{Arg}
\,\kappa$\,\cite{Cohen-Tannoudji:1973fk}. As such,
it is completely
characterized by the value of $\kappa$. In this perspective,
its classical limit as $\theta\to 0$ should be considered keeping $\kappa$ fixed.
This means that $\omega_\kappa$ tends to
the pure states $\delta_\kappa$ of $C_0(\R^2)$. Since the spectral
distance remains unchanged (it depends only on $\kappa$, not on
$\theta$), one obtains that the set of coherent states $\ccc(\omega_0)$ equipped with the
spectral distance tends to the Euclidean plane as $\theta\to 0$. The
same is true for any orbit $\ccc(\varphi)$ as soon as $\varphi$ tends to
$\delta_0$, which is for instance the case for any eigenstates of the harmonic
oscillator.

Alternatively, one can choose to characterize a 
state by the value of its
components in the eigenenergy base. For coherent states,
asking that  $c^\kappa_m$ remains constant as $\theta\to 0$ means that $\kappa$ should rescale as $\sqrt \theta$, so that
$\tilde \kappa \doteq\frac{\kappa}{\sqrt \theta}$ is unchanged.  Starting with a given a coherent state
$\omega_{\kappa_0}$ at some initial value $\theta_0$ of the deformation
parameter,  the classical limit
$\theta\to 0$ is then 
described by the net  $\omega_{\tilde\kappa_0\sqrt \theta}$ where
$\tilde\kappa_0 = \frac{\kappa_0}{\sqrt{\theta_0}}$ is a constant parameter,
fixed by the initial condition.
Whatever these initial conditions, this net tends to the origin
$\delta_0$ as $\theta\to 0$. That is, the set of coherent states tends
to a single point $\delta_0$ and $d_D(\omega_0,
\omega_{\tilde\kappa_0\sqrt \theta})= \sqrt 2 \tilde\kappa_0\sqrt
\theta$ tends to $0$ accordingly. To obtain the Euclidean plane as a limit, one
should compensate the contraction of the translation amplitude by multiplying the
spectral distance by $\theta^{-\frac 12}$. Wallet has
recently shown that such  homothetic transformations 
could be obtained by substituting the Dirac
operator $D$ with the operator 
\begin{equation}
  \label{eq:50}
  D_\omega \doteq D + \omega x_\mu\sigma^\mu \qquad \omega\in\R^{*+}
\end{equation}
introduced to study the renormalizability of the 2D Grosse-Neveu
model (see \cite{Wallet:2011uq} for details). On the whole of
$\ss(\A)$, the associated spectral distance $d_{D_\omega}$
is homothetic to the distance $d_D$, 
\begin{equation*}
  \label{eq:3ter}
  d_{D_\omega} = \frac 1{\abs{1 - \omega}} d_D.
\end{equation*}
Choosing $\omega = 1- \sqrt \theta$, one gets that $d_{D_\omega}(\omega_0,
\omega_{\tilde\kappa_0\sqrt \theta}) = \sqrt 2\tilde\kappa_o$ remains
constant, as expected.
\newline

Proposition \ref{propsymp} 
guarantees that the integral of the ``Moyal line element''
\begin{equation}
\sqrt{\omega_\kappa\circ\alpha_{R(t)}({\bf J_R}\cdot{\bf J_R})},\label{eq:6}
\end{equation}
with
 \begin{equation}
R(t) =e^{tJ_R} = \left( \begin{array}{cc} \cos t & \sin t \\ -\sin t &
     \cos t \end{array}\right), \quad
J_R = \left( \begin{array}{cc} 0 & -1 \\ 1 &
     0 \end{array}\right),
\label{eq:12}
\end{equation}
along the rotation orbit of a coherent state is greater than the
spectral distance. That is
\begin{equation}
\label{eq:52}
d_D(\omega_\kappa,\omega_\kappa\circ\alpha_{R(\tau)})\leq \int_0^\tau dt\; \sqrt{\omega_\kappa\circ\alpha_{R(t)}({\bf J_R}\cdot 
  {\bf J_R})}.
\end{equation}
It is instructive to check it explicitly. Since
\begin{equation}
  \label{eq:234}
  \omega_\kappa\circ\alpha_{R(t)}({\bf J_R}\cdot {\bf J_R}) =
  \omega_\kappa\circ\alpha_{R(t)} (x^2 + y^2) =
  2\,\omega_\kappa\circ\alpha_{R(t)}\left({\abs{z}^2}\right)= 2\, \omega_\kappa({\abs{z}^2})
\end{equation}  
as a function of $t$ is constant, the l.h.s. of (\ref{eq:52})
is the length of
the arc of angle $\tau$ and radius 
\begin{equation}
\label{eq:235} 
 r \doteq \sqrt{2\omega_\kappa({\abs{z}^2})}= \sqrt{2\omega_\kappa(\bar z\star z
+\frac{\theta}2)}  = \sqrt{2\abs{\kappa}^2 + \theta},
\end{equation}
where we use (\ref{eq:4}).
    Obviously, this length of arc is greater than the spectral distance (\ref{eq:229}) which is the length of the chord on a
circle of radius $r' =\sqrt 2\abs{\kappa}$. 

The same analysis for an arbitrary
$R(t)$-invariant state $\varphi$ yields a radius
\begin{equation}
r'= \sqrt{\abs{2\kappa}^2
  + \theta +2\varphi(\bar z\star z)}
\label{eq:8}
\end{equation}
while $r$ remain unchanged. Interestingly, the difference
$r' -r$ between the two radius  is minimal
for coherent states ($\omega_0(\bar z\star z) =0)$ and vanishes in the classical limit $\theta \to 0$.

\subsection{Quantum length in the DFR model}
\label{secDFR}
We summarize in this section the analysis developed at length
in \cite{Martinetti:2011fk}. The $2N$-dimensional DFR model of quantum spacetime (see
\cite{Doplicher:2001fk, Doplicher:2006uq} as well as
\cite{Piacitelli:2010uq} for a recent
 review) is described by
coordinate operators $q_\mu$, $\mu=1, 2N$, that satisfy the commutation
relations \cite{Doplicher:1995hc}
 \begin{equation}
  \label{eq:61}
  [q_\mu, q_\nu] = i\lambda_P^2 \Theta_{\mu\nu}\ii,
\end{equation}
with $\Theta$ the matrix given in (\ref{eq:45}) and $\lambda_P$ the
Planck length. It carries a
representation of the Poincar\'e group $G$ under which (\ref{eq:61}) is
covariant (the left-hand side transforms under $\text{Ad}\,G$). We 
shall not take into account this
action here, since we are interested in the \emph{Euclidean} length operator, 
\begin{equation}
  \label{eq:115}
  L\doteq \sqrt{\sum_{\mu=1}^{2N} dq_\mu^2}, \quad\quad dq_\mu \doteq q_\mu\otimes \ii - \ii\otimes q_\mu,
\end{equation}
whose spectrum is obviously not
Poincar\'e invariant. Said differently, we fix
once and for all the matrix $\Theta$ in (\ref{eq:61}). Incidentally, this
means that our analysis also applies to the so-called canonical
noncommutative spacetime (or $\theta$-Minkowski), characterized by the
invariance (opposed to covariance) of the commutators (\ref{eq:61})
under the action of the quantum group
$\theta$-Poincar\'e. In both models, the length operator $L$ is
promoted to a quantum observable \cite{Amelino-Camelia:2009fk, Bahns:2010fk}, and
\begin{equation}
  \label{eq:119}
  l_p \doteq \text{min} \{\lambda \in \text{Sp } L\}
\end{equation}
is interpreted as the minimal value that may come out from a length
measurement. 

The link with the spectral distance is obtained by identifying $q_\mu$
with the Moyal coordinate $x_\mu$, viewed as an unbounded operator
affiliated \cite{Woronowicz:1991fk} to $\kk$. The choice of the representation, left-regular on 
$\hh=L^2(\R^{2N})$, that is $q_\mu= \ll(x_\mu)$; or integrated Schr\"odinger on $L^2(\R^N)$, $q_\mu= \pi_S(x_\mu)$; is not
relevant for the following discussion. In both cases, $q_\mu$ is
affiliated to $\kk$. To any pair of states
$(\varphi,\tilde\varphi)\in\ss(\kk)\times\ss(\kk)$ in the domain of the $q_\mu$'s,
one associates the \emph{quantum length}
\cite{Martinetti:2011fk}
\begin{equation}
  \label{eq:116}
  d_{L}(\varphi, \tilde\varphi) \doteq (\varphi\otimes\tilde\varphi)(L).
\end{equation}
Obviously $d_L$ is not a distance: for $N=1$, an explicit computation yields
\begin{equation}
  \label{eq:121}
  l_p = \sqrt 2 \lambda_P,
\end{equation}
so that $d_L(\varphi,\varphi)\geq l_p$ never vanishes. Consequently, there is a priori
little sense to
compare the quantum length  with the spectral distance. 

Nevertheless, 
we have shown in \cite{Martinetti:2011fk} that it does make sense to compare
the quantum square-length,
\begin{equation}
  \label{eq:1160}
  d_{L^2}(\varphi, \tilde\varphi) \doteq (\varphi\otimes\tilde\varphi)(L^2),
\end{equation} 
with the (square of the) spectral distance $d_{D'}$ in the
double unital Moyal space of section \ref{secPyth}. The doubling procedure
allows to implement the notion of minimal length between a state and itself into
the spectral distance framework, by identifying $d_{L^2}(\varphi, \varphi)$ to
$d^2_{D'}(\varphi^1, \varphi^2)$. Technically, this simply  amounts to fix the parameter $\Lambda$ in the Dirac
operator $D_2$ to the required value
since, by (\ref{distint}), one has
\begin{equation}
d_{L^2}(\varphi, \varphi) = d^2_{D'}(\varphi^1,
\varphi^2)\quad \text{ if and only if } \quad\Lambda =
d_{L^2}(\varphi, \varphi)^{-\frac 12}.
\label{eq:166}
\end{equation}

Since $d_{D'}(\varphi, \varphi)$, as a function of $\varphi$, is
constantly equal to $\Lambda^{-1}$ on $\sa$ (in fact, one could make it non-constant by
introducing a Higgs field; this point is discussed in
\cite{Martinetti:2011fk}), once the free parameter $\Lambda$ is fixed,
the identification of $d^2_{D'}$ with
$d_{L^2}$ may make sense only for those states $\tilde\varphi$ such
that $d_{L^2}(\tilde\varphi, \tilde\varphi) = d_{L^2}(\varphi,
\varphi)$. This is indeed the case for the states in the set
$\ccc(\varphi)$ defined in (\ref{eq:162}). One then gets by theorem
\ref{theopyth}, that the identification $d_{L^2} \leftrightarrow d^2_{D'}$
extends to any pair of states $(\varphi, \tilde\varphi)$ with $\ccc(\varphi)\ni\tilde\varphi\neq\varphi$ 
if and only if the spectral distance on a 
\emph{single copy} of
the Moyal plane is
\begin{equation}
  \label{eq:117}
  d_D(\varphi,\tilde\varphi) = \sqrt{d_{L^2}(\varphi,\tilde\varphi) -d_{L^2}(\varphi,\varphi)}.
\end{equation}

Eq. (\ref{eq:117}) is the true condition guaranteeing that, once
the obvious discrepancy due to the non vanishing of
$d_{L^2}(\varphi, \varphi)$ is solved, the spectral distance and the
quantum length capture the same metric information on the Moyal plane.
Remarkably, this conditions holds true for the states that are of particular physical importance from the DFR
point of view,  namely the states $\varphi$ of \emph{optimal
  localization},
for which
\begin{equation}
  \label{eq:169}
  d_{L^2}(\varphi, \varphi) = l_P.
\end{equation}
Indeed, these states are nothing but the coherent states discussed in the previous
subsection. By proposition \ref{propcoherent0} and the explicit
computation of $d_{L^2}(\varphi, \tilde\varphi)$ carried out in
\cite{Martinetti:2011fk}, one shows that (\ref{eq:117}) actually holds
true.

\section*{VII \hspace{.35truecm}Conclusion}

We have proved two theorems on Connes' spectral
distance on the Moyal plane: first, the spectral distance between any state of the
Moyal algebra and any of its translated is precisely the amplitude of
the translation; second the product of the Moyal plane by
$\cc^2$ is an orthogonal in the sense of Pythagoras theorem, restricted to
the orbit of the translation group in the space of states.
These results allow to identify the spectral distance with the DFR quantum length between states
of optimal localization.

We discussed the extension of these results to other orbits, and showed that the techniques developed for
translations do not directly apply to symplectic transformations.
 Another tempting generalization is to consider, in the
spirit of Rieffel deformation,  the spectral
distance associated to the deformation of an algebra (not necessarily
the Schwartz functions) by the action of a Lie group $G$ (not necessarily
$\R^2$) with a Dirac operator given by the action of
some elements of $\frak g = \text{Lie } G$. One could expect the
distance on the orbit of $e^{it\frak g}$ to be again the amplitude of
translation. This will be the object of
future work.  
\begin{center}
\rule{5cm}{.7pt}
\end{center}

\section*{Acknowledgments}
Warm thanks to Francesco D'Andrea and Fedele Lizzi for various
discussions, remarks and suggestions on this work, as well as for their
fair-play.  Many thanks to  Jean-Christophe Wallet for a
careful reading of the manuscript and suggestions, and to  Sergio
Doplicher for several illuminating discussions.

Work supported by an ERG-Marie
Curie fellowship 237927 \emph{Noncommutative geometry and quantum
gravity} and by the ERC Advanced Grant 227458 OACFT \emph{Operator Algebras and
Conformal Field Theory}.
\newpage

\small{\bibliographystyle{abbrv}

\begin{thebibliography}{10}

\bibitem{Amelino-Camelia:2009fk}
G.~Amelino-Camelia, G.~Gubitosi, and F.~Mercati.
\newblock Discretness of area in noncommutative space.
\newblock {\em Phys. Lett. B} 676:180--83, 2009.

\bibitem{Collaboration:2012fk}
The Atlas Collaboration.
\newblock Observation of a new particle in the search for the standard model
  Higgs boson with the ATLAS detector at the LHC, \emph{arXiv:1207.7214 [hep-ex]} 2012.
The CMS. collaboration.
\newblock Observation of a new boson at mass of 125GeV with the CMS experiment
  at the LHC.
\emph{arXiv:1207.7235 [hep-ex]} 2012.


\bibitem{Bahns:2010fk}
D.~Bahns, S.~Doplicher, K.~Fredenhagen, and G.~Piacitelli.
\newblock Quantum geometry on quantum spacetime: distance, area and volume
  operators.
\newblock {\em Commun. Math. Phys.} 308:567--589, 2011.

\bibitem{Bellissard:2010fk}
J.~V. Bellissard, M.~Marcolli, and K.~Reihani.
\newblock Dynamical systems on spectral metric spaces.
\newblock {\em arXiv:1008.4617v1 [math.OA]}, 2010.

\bibitem{bimonte}
G.~Bimonte, F.~Lizzi, and G.~Sparano.
\newblock Distances on a lattice from noncommutative geometry.
\newblock {\em Phys. Lett. B} 341:139--146, 1994.

\bibitem{Bondia:1988nr}
J.~M.~G. Bondia and J.~C. Varilly.
\newblock Algebras of distributions suitable for phase-space quantum
mechanics  {I}.
\newblock {\em J. Math. Phys.} 29(4):869--879, 1988.

\bibitem{Bondia:1988qv}
J.~M.~G. Bondia and J.~C. Varilly.
\newblock Algebras of distributions suitable for phase-space quantum mechanics.
  {II}.
\newblock {\em J. Math. Phys.}, 29(4):880--887, 1988.

\bibitem{Cagnache:2009oe}
E.~Cagnache, F.~d'Andrea, P.~Martinetti, and J.-C. Wallet.
\newblock The spectral distance on {M}oyal plane.
\newblock {\em J. Geom. Phys.} 61:1881--1897, 2011.

\bibitem{Cagnache:2009vn}
E.~Cagnache and J.-C. Wallet.
\newblock Spectral distances: Results for Moyal plane and noncommutative torus.
\newblock {\em SIGMA} 6(026):17 pages, 2010.

\bibitem{Chamseddine:2007oz}
A.~H. Chamseddine, A.~Connes, and M.~Marcolli.
\newblock Gravity and the standard model with neutrino mixing.
\newblock {\em Adv. Theor. Math. Phys.} 11:991--1089, 2007.

\bibitem{Christensen:2006fk}
E.~Christensen and C.~Ivan.
\newblock Spectral triples for af {C}*-algebras and metrics on the cantor set.
\newblock {\em J. Operator Theory} 56(1):17--46, 2006.

\bibitem{Christensen:2011fk}
E.~Christensen, C.~Ivan, and E.~Schrohe.
\newblock Spectral triples and the geometry of fractals.
\newblock {\em Journal of Noncommutative Geometry}, 2011.

\bibitem{Cohen-Tannoudji:1973fk}
C.~Cohen-Tannoudji, B.~Diu, and F.~Lalo\"e.
\newblock {\em M\'ecanique quantique I}.
\newblock Hermann, Paris, 1973.

\bibitem{Connes:1989fk}
A.~Connes.
\newblock Compact metric spaces, Fredholm modules, and hyperfiniteness.
\newblock {\em Ergod. Th. \& Dynam. Sys.} 9:207--220, 1989.

\bibitem{Connes:1994kx}
A.~Connes.
\newblock {\em Noncommutative Geometry}.
\newblock Academic Press, 1994.

\bibitem{Connes:1996fu}
A.~Connes.
\newblock Gravity coupled with matter and the foundations of noncommutative
  geometry.
\newblock {\em Commun. Math. Phys.} 182:155--176, 1996.

\bibitem{Connes:2002xr}
A.~Connes and M.~Dubois-Violette.
\newblock Noncommutative finite-dimensional manifolds {I}. spherical manifolds
  and related examples.
\newblock {\em Commun. Math. Phys.} 230:539--579, 2002.

\bibitem{pekin}
J.~Dai and X.~Song.
\newblock {P}ythagoras' theorem on a 2d-lattice from a "natural" {D}irac
  operator and {C}onnes' distance formula.
\newblock {\em J. Phys. A} 34:5571--5582, 2001.

\bibitem{dAndrea:2009xr}
F.~D'Andrea and P.~Martinetti.
\newblock A view on optimal transport from noncommutative geometry.
\newblock {\em SIGMA} 6(057):24 pages, 2010.

\bibitem{DAndrea:2012fk}
F.~D'Andrea and P.~Martinetti.
\newblock On {P}ythagoras theorem for products of spectral triples.
\newblock {To be published in Lett. Math. Phys.} {\em arXiv:1203.3184 [math-ph]}, 2012.

\bibitem{dimakis2}
A.~Dimakis and F.~Mueller-Hoissen.
\newblock {C}onnes' distance function on one dimensional lattices.
\newblock {\em Int. J. Theor. Phys.} 37:907, 1998.

\bibitem{Doplicher:2001fk}
S.~Doplicher.
\newblock Spacetime and fields, a quantum texture.
\newblock {\em Proceedings 37th Karpacz Winter School of Theo. Physics}, pages
  204--213, 2001.

\bibitem{Doplicher:2006uq}
S.~Doplicher.
\newblock Quantum field theory on quantum spacetime.
\newblock {\em J. Phys.: Conf. Ser.} 53:793--798, 2006.

\bibitem{Doplicher:1995hc}
S.~Doplicher, K.~Fredenhagen, and J.~E. Robert.
\newblock The quantum structure of spacetime at the {P}lanck scale and quantum
  fields.
\newblock {\em Commun. Math. Phys.} 172:187--220, 1995.

\bibitem{Figueroa:2001fk}
H.~Figueroa, J.~M.~G. Bondia, and J.~C. Varilly.
\newblock {\em Elements of Noncommutative Geometry}.
\newblock Birkhauser, 2001.

\bibitem{Gayral:2004rc}
V.~Gayral, J.~M.~G. Bondia, B.~Iochum, T.~Sch{\"u}cker, and J.~C. Varilly.
\newblock Moyal planes are spectral triples.
\newblock {\em Commun. Math. Phys.} 246:569--623, 2004.

\bibitem{GGBNV} M. Gadella, J.M. Gracia-Bondia, L.M. Nieto and
  J.C. Varilly.
\newblock Quadratic Hamiltonians in phase-space quantum mechanics.
\newblock{\em  J. Phys. A} 22:2709--2738,1989.

\bibitem{Groene}
H.~Groenewold.
\newblock {O}n the principles of elementary quantum mechanics.
\newblock {\em Physica} 12:405--460, 1946.

\bibitem{Gromov:1999fk}
M.~Gromov.
\newblock {\em Metric structures for Riemannian and non-Riemannian spaces}, Birkh{\"a}user 1999.

\bibitem{Iochum:2001fv}
B.~Iochum, T.~Krajewski, and P.~Martinetti.
\newblock Distances in finite spaces from noncommutative geometry.
\newblock {\em J. Geom. Phy.} 31:100--125, 2001.

\bibitem{Kadison19861}
R.~V. Kadison and J.~R. Ringrose.
\newblock {\em Fundamentals of the Theory of Operator Algebras. {V}olume {I},
  {A}dvanced theory}.
\newblock Academic Press, 1986.

\bibitem{Kadison1986}
R.~V. Kadison and J.~R. Ringrose.
\newblock {\em Fundamentals of the Theory of Operator Algebras. {V}olume {II},
  {A}dvanced theory}.
\newblock Academic Press, 1986.

\bibitem{lizzif}
F.~Lizzi, J.~Varilly, and A.~Zamponi.
\newblock {P}rivate communication.

\bibitem{Lizzi:2001fk}
F.~Lizzi, R.~J. Szabo, and A.~Zampini.
\newblock Geometry of the gauge algebra in noncommutative Yang-Mills theory.
\newblock {\em JHEP} 0108(032), 2001.

\bibitem{madore}
J.~Madore.
\newblock {\em An introduction to noncommutative differential geometry and its
  physical applications}.
\newblock Cambridge University Press, 1995.

\bibitem{Martinetti:2001fk}
P.~Martinetti.
\newblock Distances en g{\'e}om{\'e}trie non-commutative.
PhD thesis, \newblock {\em arXiv:math-ph/0112038}, 2001.

\bibitem{Martinetti:2006db}
P.~Martinetti.
\newblock {C}arnot-{C}arath{\'e}odory metric and gauge fluctuation in
  noncommutative geometry.
\newblock {\em Commun. Math. Phys.} 265:585--616, 2006.

\bibitem{Martinetti:2006fk}
P.~Martinetti.
\newblock {C}arnot-{C}arath{\'e}odory metric vs gauge fluctuation in
  noncommutative geometry.
\newblock {\em African Journal of Mathematical Physics} 3:157--162, 2006.

\bibitem{Martinetti:2008hl}
P.~Martinetti.
\newblock Spectral distance on the circle.
\newblock {\em J. Func. Anal.} 255:1575--1612, 2008.

\bibitem{Martinetti:2006rz}
P.~Martinetti.
\newblock Smoother than a circle or how noncommutative geometry provides the
  torus with an egocentric metric.
\newblock In C.~university~press (Roumania), editor, {\em proceedings of Deva
  intl. conf. on differential geometry and physics}, 2006.

\bibitem{Martinetti:2011fkbis}
P.~Martinetti and L.~Tomassini.
\newblock Length and distance on a quantum space.
\newblock {\em Proc. of Sciences}, 042, 2011.

\bibitem{Tomarsinetti}
P.~Martinetti and L.~Tomassini, in progress.

\bibitem{Martinetti:2011fk}
P.~Martinetti, F.~Mercati, and L.~Tomassini.
\newblock Minimal length in quantum space and integrations of the line element
  in noncommutative geometry.
\newblock {\em Rev. Math. Phys.} 24(5):36 pages, 2012.

\bibitem{Martinetti:2002ij}
P.~Martinetti and R.~Wulkenhaar.
\newblock Discrete {K}aluza-{K}lein from scalar fluctuations in noncommutative
  geometry.
\newblock {\em J. Math. Phys.} 43(1):182--204, 2002.

\bibitem{Moyal}
J.~E. Moyal.
\newblock Quantum mechanics as a statistical theory.
\newblock {\em Proc. Camb. Phil. Soc.} 45:99--124, 1949.

\bibitem{Piacitelli:2010uq}
G.~Piacitelli.
\newblock Quantum spacetime: a disambiguation.
\newblock {\em SIGMA}  6(073):43 pages, 2010.
\bibitem{Reed1975}
M.~Reed and B.~Simon.
\newblock {\em Methods of modern mathematical physics. Vol 2. Fourier analysis,
  self-adjointness.}
\newblock AP, 1975.

\bibitem{Rieffel:1999ec}
M.~A. Rieffel.
\newblock Metric on state spaces.
\newblock {\em Documenta Math.} 4:559--600, 1999.

\bibitem{Sakai1971}
S.~Sakai.
\newblock {\em {C}$^*$-algebras and {W}$^*$-algebras}.
\newblock Springer-Verlag, Berlin, 1971.

\bibitem{sitarz}
A.~Sitarz.
\newblock Rieffel deformation quantization and isospectral deformation.
\newblock {\em Int. J. Theor. Phys.} 40:1693--1696, 2001.

\bibitem{Rudin:1970fk}
W.~Rudin.
\emph{Real and complex analysis}, McGraw-Hill, 1970.

\bibitem{Villani:2003fk}
C.~Villani.
\newblock {\em Topics in Optimal Transportation}, volume~58.
\newblock 2003.

\bibitem{Wallet:2011uq}
J.-C. Wallet.
\newblock Connes distance by examples: Homothetic spectral metric spaces.
\newblock {\em Rev. Math. Phys.} 24(9):26 pages, 2012.

\bibitem{Woronowicz:1991fk}
S.~L. Woronowicz.
\newblock Unbounded elements affiliated with {C}*-algebras and non-compact
  quantum groups.
\newblock {\em Commun. Math. Phys.} 136:399-432, 1991.
\end{thebibliography}

\end{document}